\documentclass{aa}  
\usepackage{graphicx}
\usepackage{txfonts}
\usepackage{epsfig}
\usepackage{times}
\usepackage{natbib}
\usepackage{url}
\usepackage{color}
\usepackage{amssymb,amsmath}
\usepackage{enumerate}
\usepackage[bookmarks = true,colorlinks = true,linkcolor = blue,urlcolor = blue,citecolor = blue,bookmarks = true,linktocpage = true,hyperindex = true,breaklinks]{hyperref}

\graphicspath{ {./figures/} }

\begin{document}
\title{
Multi thermal 
atmosphere of
 a mini 
 solar flare during magnetic reconnection observed with IRIS
 }

\author{Reetika Joshi
\inst{1,2}
Brigitte Schmieder
\inst{1,3,4},
Akiko Tei
\inst{5},
Guillaume Aulanier
\inst{1},
Juraj L\"{o}rin\v{c}\'{i}k
\inst{6,7}
Ramesh Chandra
\inst{2},
\and
Petr Heinzel
\inst{6}
}
\institute{LESIA, Observatoire de Paris, Universit\'e PSL, CNRS, Sorbonne Universit\'e,
      Universit\'e de Paris, 5 place Jules Janssen, F-92190 Meudon, France
\and
Department of Physics, DSB Campus, Kumaun University, Nainital -- 263 001, India\\
\email{reetikajoshi.ntl@gmail.com, reetika.joshi@obspm.fr}
\and
Centre for Mathematical Plasma Astrophysics, Dept. of Mathematics, KU Leuven, 3001 Leuven, Belgium
 \and
University of Glasgow, School of Physics and Astronomy, Glasgow, G128QQ, Scotland
\and
Institute of Space and Astronautical Science, Japan Aerospace Exploration Agency, 3-1-1 Yoshinodai, Chuo-ku, Sagamihara, Kanagawa 252-5210, Japan
\and
Astronomical Institute of the Czech Academy of Sciences, Fri\v{c}ova 298, 251 65 Ond\v{r}ejov, Czech Republic
\and
Institute of Astronomy, Charles University, V Hole\v{s}ovi\v{c}k\'{a}ch 2, CZ-18000 Prague 8, Czech Republic
}
\authorrunning{Reetika Joshi et al.} 
   \titlerunning{Stratification of the multi thermal solar atmosphere}
\abstract
{\emph{The Interface Region Imaging Spectrograph} (IRIS) with its high spatial and temporal resolution brings exceptional plasma diagnostics of 
solar chromospheric and coronal activity during magnetic reconnection.
}
{The aim of this work is to study 
the fine structure and dynamics of the plasma at a jet base  forming   a mini flare 
between two emerging magnetic fluxes (EMFs)
observed with IRIS and the \emph{Solar Dynamics Observatory} (SDO) instruments.
}
{We proceed to a spatio-temporal analysis of IRIS spectra observed in the spectral ranges of Mg II, C II and Si IV ions. Doppler velocities from Mg II lines are computed by using a cloud model technique.
}
 {Strong asymmetric Mg II and C II line profiles with extended blue wings observed at the reconnection site (jet base) are interpreted by the presence 
 of two chromospheric temperature clouds, one explosive cloud with  blueshifts 
 at 
 290 km s$^{-1}$ 
 and one cloud with smaller Dopplershift (around 36 km s$^{-1}$).
Simultaneously at the same location (jet base), strong  emission of several transition region lines (e.g. 
O IV and Si IV), 
 emission of the Mg II triplet lines of the Balmer-continuum and 
  absorption  of  identified chromospheric lines in  Si IV broad profiles 
 have been 
 observed and analysed.
}
{
Such observations of 
IRIS line and continuum  emissions allow  us to propose a stratification 
model for the white-light mini flare atmosphere with multiple layers of different temperatures along the line of sight, 
in a reconnection current sheet. It is the first time that we could quantify the fast speed (possibly Alfv\'enic flows) of cool clouds  ejected perpendicularly to the jet direction  by using the cloud model technique. We conjecture that 
the 
ejected clouds
come from   plasma  
which was trapped between the two EMFs before 
 reconnection
 or be caused by 
 chromospheric-temperature (cool) upflow  material like  in a surge, 
 during reconnection.
}
\maketitle \section{Introduction} 
\label{intro}
The \emph{ Interface Region Imaging Spectrograph} \citep[IRIS,] [] {Pontieu2014} has revealed several transient small scale  phenomena 
in the solar atmosphere  
such as 
UV bursts (see the review of \citet{Young2018}) recently called IRIS bombs or IBs \citep{Peter2014,Grubecka2016,Chitta2017,Tian2018}
explosive events \citep{Kim2015,Chen2019,Gupta2015,Huang2017,Ruan2019}, blow jets \citep{Shen2017} 
 and bidirectional outflow jets \citep{Ruan2019}. 
UV bursts are very tiny bright points with a  bright core less than 2 arcsec. 
Their lifetime is short ($\approx$ 10 sec) but with possibly recurrent enhancements during one hour giving the impression of flickering \citep{Pariat2007}.  
 
 With IRIS, UV bursts are observed in chromospheric lines  with extended wings (Mg II and C II), in transition region temperature line (Si IV) but have no signature in coronal lines.
 Si IV line profiles in UV bursts are  commonly very wide, over 2.5 \AA\  \citep{Vissers2015}. In the IB observations of \citet{Peter2014} the two Si IV line wings presented a peak  
 at $\pm$ {200 km s$^{-1}$} separated from the line center  with  intensity  enhanced by a factor of 1000 compared to the surrounding atmosphere. These two peaks suggested bilateral outflows.
  In  such broad Si IV profiles, dips corresponding to chromospheric temperature formation lines, e.g.  Ni II at 1393.33 \AA\ were observed, indicating the presence of  cool plasma  (10$^4$ K)
along the line of sight (LOS) \citep{Peter2014}. 
 From these  spectral observations  \citet{Peter2014}
 concluded 
 that hot pockets (100,000 K)  were present in the photosphere.
 10 to 20 \% of UV  brightenings are related to  Ellerman bombs (EBs) characterized by  the emission of  far wing extension in chromospheric  line profiles  (H$\alpha$,  Mg II).  The question of formation heights of IBs and EBs arised \citep{Grubecka2016}.  \citet{Hansteen2019} unified  the problem by proposing numerical MHD simulations based on the  1.5D RH code \citep{Pereira2015}
 and the fully MULTI3D code \citet{Leenaarts2009}
 and found that
 IBs and EBs  correspond to the same  reconnection event,  the  reconnection occurring in different altitudes along the same vertical current sheet from the deep chromosphere to the corona. Si IV synthetized lines matched well with   broad Si IV profiles observed in  IBs  (\citep{Peter2014}). An other attempt  to understand the height formation of EBs and IBs has been made by 
 \citet{Grubecka2016} where  the NLTE radiative transfer code  in a 1D atmosphere model of \citet{Berlicki2014} has been developed  for the Mg II element.  \citet{Grubecka2016} could fit  the  Mg II h and k line profiles  in IBs and EBs by a deposit of heating  at different levels in the atmosphere,  between the photosphere (50 km)  and high chromosphere (900 km). Their 1 D model is valid until the ionisation degree temperature of Mg II, so that they  could not compute synthesized Si IV line profiles. More recently using  the RADYN simulation code combined with the MULTI3D code, \citet{Reid2017} obtained synthesized line profiles for three elements: Mg II, Ca II, hydrogen (H$\alpha$) in  Ellerman bombs  with a deposit of energy at different altitudes   between 300 and 1000 km in their 1D model.
 However they could neither work on the Si IV lines for the same reasons as \citet{Grubecka2016},  neither fit the three lines of the three different elements  simultaneously. A complete different point of view was brought by \citet{Judge2015}  where it was proposed that IB spectra  shape is due to 
 Alfv\'enic turbulence.
 \begin{figure*}
\hspace{1.4cm}
\centering
\includegraphics[width=1.0
\textwidth]{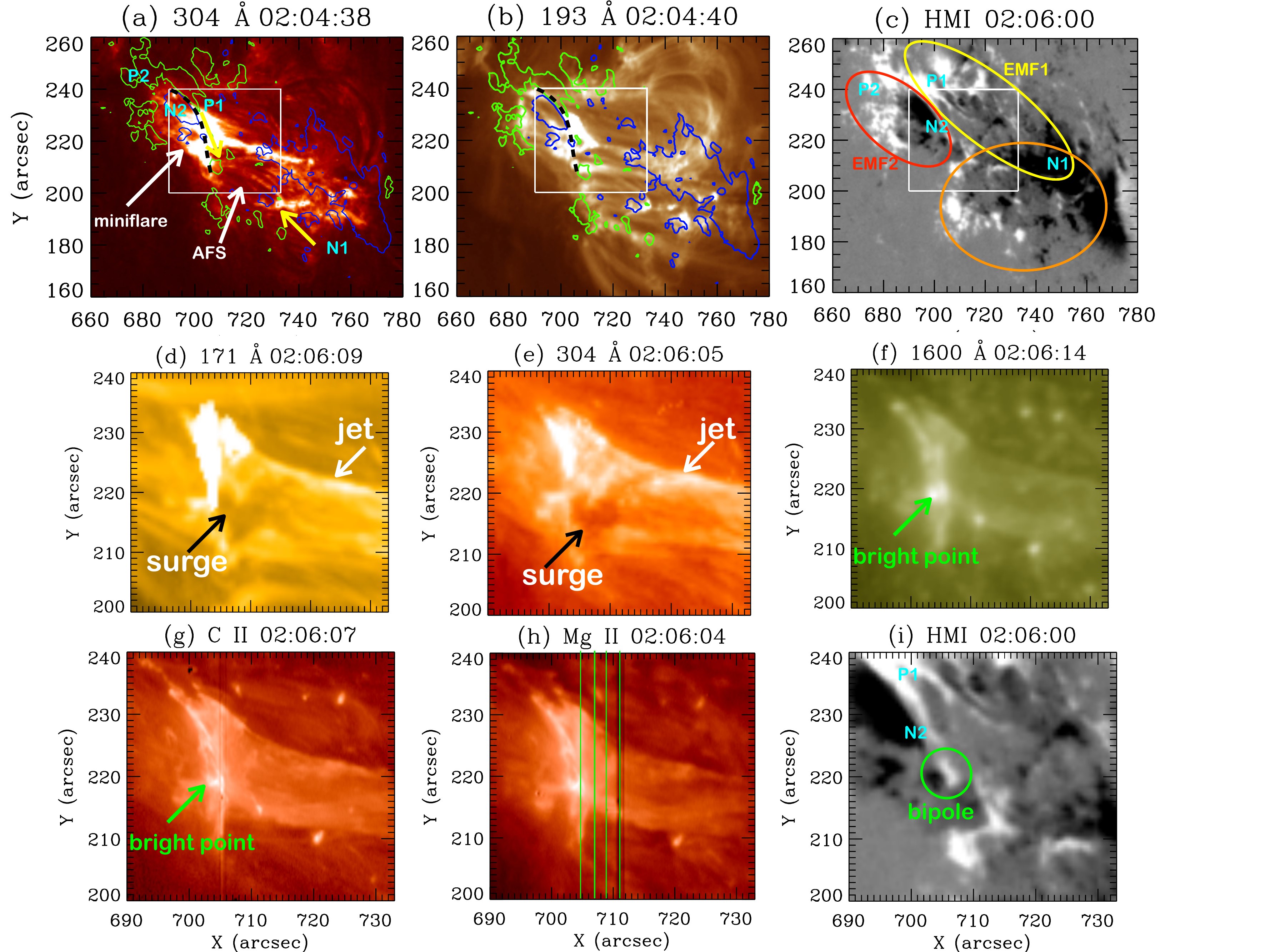}
\caption{Multi-wavelength observations of the  solar jet and surge in different AIA and IRIS wavebands on March 22, 2019. Panels (a-b) show the active region AR 12736  with the mini flare in  AIA 304 \AA\ and 193 \AA\ passbands respectively. Panel (c) shows the longitudinal magnetic field configuration observed with HMI consisting of emerging magnetic fields (EMF): EMF1 (P1-N1)  and EMF2 (P2-N2) and an earlier EMF encircled respectively by yellow, red, orange ovals. 
 On panels (a-b) the maps are overlaid with the HMI magnetic field contours of strength $\pm$ 300 Gauss, where green contours are for positive and blue  for negative magnetic polarity, 
 the magnetic polarity inversion line (PIL) between P1 and N2
along the North-South bright area  is indicated by a dashed black line and  the  dark elongated structures between N1 (yellow arrow)  and P1
(AFS), are pointed by a white arrow. 
The small white square in panels (a-c) shows the FOV for the  panels (d to i).
Panels (d-f) represent three different AIA channels (171 \AA, 304 \AA, and 1600 \AA), zooming on the jet base in triangle-shape and the surge
indicated with white and black arrows respectively. 
Panels (g-h) are IRIS SJI 1330 \AA\, and SJI 2796 \AA.
The four positions of  the IRIS slit are shown with green vertical lines in panel (h).
Panel (i) is a zoom  of the magnetic  configuration 
at the jet base along the PIL  between P1 and N2. The polarities P1 and N2 are very much extended along the PIL between them. N2 is continuously sliding along (see magnetic field evolution in an attached animation).
In panel (i), the bipole  (part of N2-P1) where the reconnection takes place
is encircled with green color. This region corresponds  to the bright point indicated by a green arrow in panels (f and g). North is up and West is in the right in all the panels.}
\label{AIA_IRIS}
\end{figure*}
Broadened Si IV profiles  could be also due to the sum of different structures having rapidly changing velocity  however it  does not seem to  be the admitted solution for IBs because they are relatively stable during time, {\it e.g.} 15 minutes in the UV bursts studied by \citet{Gupta2015}. 
For other observations of UV bursts the dip in double peaked Si IV line profiles is  interpreted to be caused by self absorption mechanism
\citep{Yan2015}.
Si IV profiles observed in UV bursts vary spatially significantly 
across the IBs \citep{Yan2015,Grubecka2016,Chitta2017}.  

With IRIS  the chromospheric C II and Mg II lines are frequently observed not only in the UV bursts but also in the quiet chromosphere as well as in solar flares and jets \citep{Leenaarts2013a,Rathore2015}. They  are optically--thick 
  lines 
  and need  a radiative transfer approach to determine the physical quantities of  plasma.
  The Mg II h and k resonance lines in the quiet Sun are formed over a wide range of chromospheric heights. They usually appear as doubly peaked profiles with a central reversal. 
Simulations in the  quiet chromosphere has been carried out by \citet{Athay1968, Milkey1974, Ayres1976,Uitenbroek1997,Lemaire2004,Leenaarts2013a,Leenaarts2013b,Pereira2013}.
The core of the line is formed just under the transition region (T$<$ 20,000K), the wings at the minimum of temperature (T = 5000 K). 

IRIS  spectral data allow to make many  progresses on the plasma diagnostics in flares.
\citet{Kerr2015} and \citet{Liu2015}
 recently discussed the emission of chromospheric  lines as observed in solar flares. They said about these lines that: "They appeared as redshifted, single-peaked profiles, however some pixels present 
 a net blue asymmetry".  
 The blue asymmetry can be explained by down-flowing plasma absorbing the red peak emission and not by strong blueshift emission 
 \citep{Berlicki2005}.
\begin{figure*}[ht!]
\includegraphics[width=1.0
\textwidth]{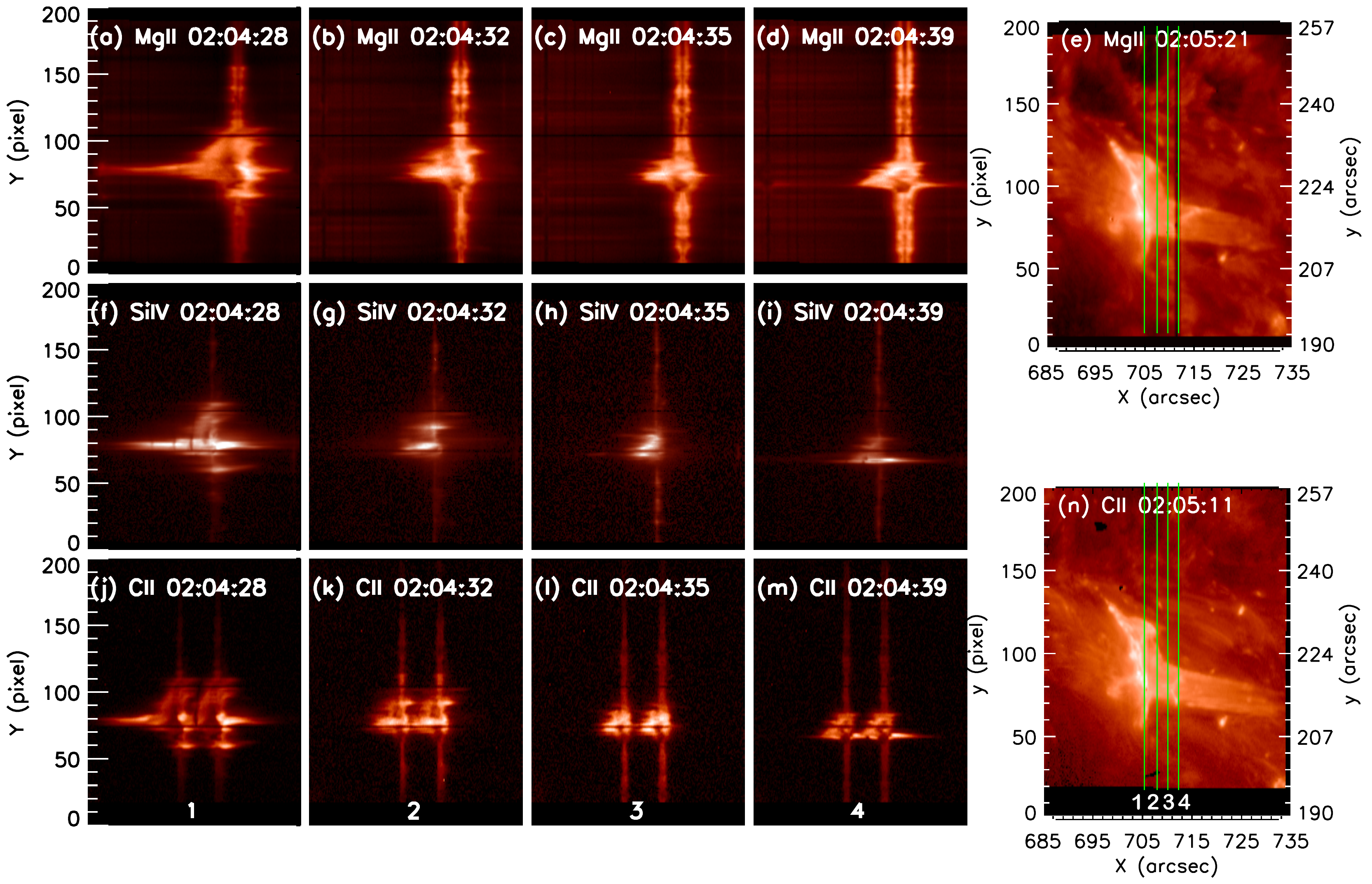}
\caption{({\it Left columns 1- 4}) IRIS spectra of the Mg II k line at 2796.35 \AA\ (a-d), Si IV 1402.77 \AA\ line (f-i), and   C II 1330 \AA\ line (j-m) at the four slit positions of raster 89. ({\it from left to right}), slit positions 1, 2, 3, 4 are corresponding to  spectra  356, 357,358, 359. 
 The  spectra in slit position 1 ({\it column} 1) corresponds to the region of the 
mini flare at the reconnection site. ({\it  Right column}) SJI 2796  
(panel e)  and  SJI 1330 
(panel n) 
with   four green solid  vertical lines  indicating  the  four  slit positions of the raster. The y axis  unit  in the SJIs (e and n) is  in pixels on the left and arcsec on the right to show the correspondence between spectra and SJIs.
} \label{spectra}
\end{figure*}

Chromospheric response to intense heating, even in the 1D model, is complicated. The shape of the emission line profiles depends sensitively on the physical conditions of the plasma and its dynamics, in particular the plasma flows that arise at the line core formation heights. 
They may have symmetrical profiles.
Moreover the highest Near Ultraviolet (NUV) continuum enhancements  observed in strong flares are most
likely because of the Balmer continuum  formed by Hydrogen recombination \citep{Kleint2017} and consequently  flares can be assimilated to  white light flares, commonly observed in optical continuum  where the energy  deposit is localized at  the minimum temperature  region.

The ratio of  IRIS transition  region lines is also a good diagnostics for the determination of  the plasma density in flares \citep{Polito2016,Dudik2017}.  However theoretical simulations showed that this analysis is valid only if there is no self absorption in   the transition lines like Si IV  and/or if  the Si IV lines are not optically thick \citep{Dudik2017,Kerr2019}.
The electron density ({\it N}{$_e$}) in flare ribbons can be enhanced by two orders of magnitude more than in plage region ({\it N}{$_e$} $>$ 10 $^{13}$ cm$^{-3}$). 
For multiple flaring kernels, chromospheric  lines show a rapidly evolving
double-component structure: an enhanced emission component at rest, and a broad, highly red-shifted
component of comparable intensity. \citet{Graham2020} interpreted such observations by beams penetrating very deep in the atmosphere. The red-shifted components migrate from redshifts towards
the rest wavelength within 30 seconds. The electron beams would  dissipate their
energy higher, driving an explosive evaporation, and a counterpart condensation is created as a very dense layer.

\begin{figure*}[ht!]
\includegraphics[width=1.0
\textwidth]{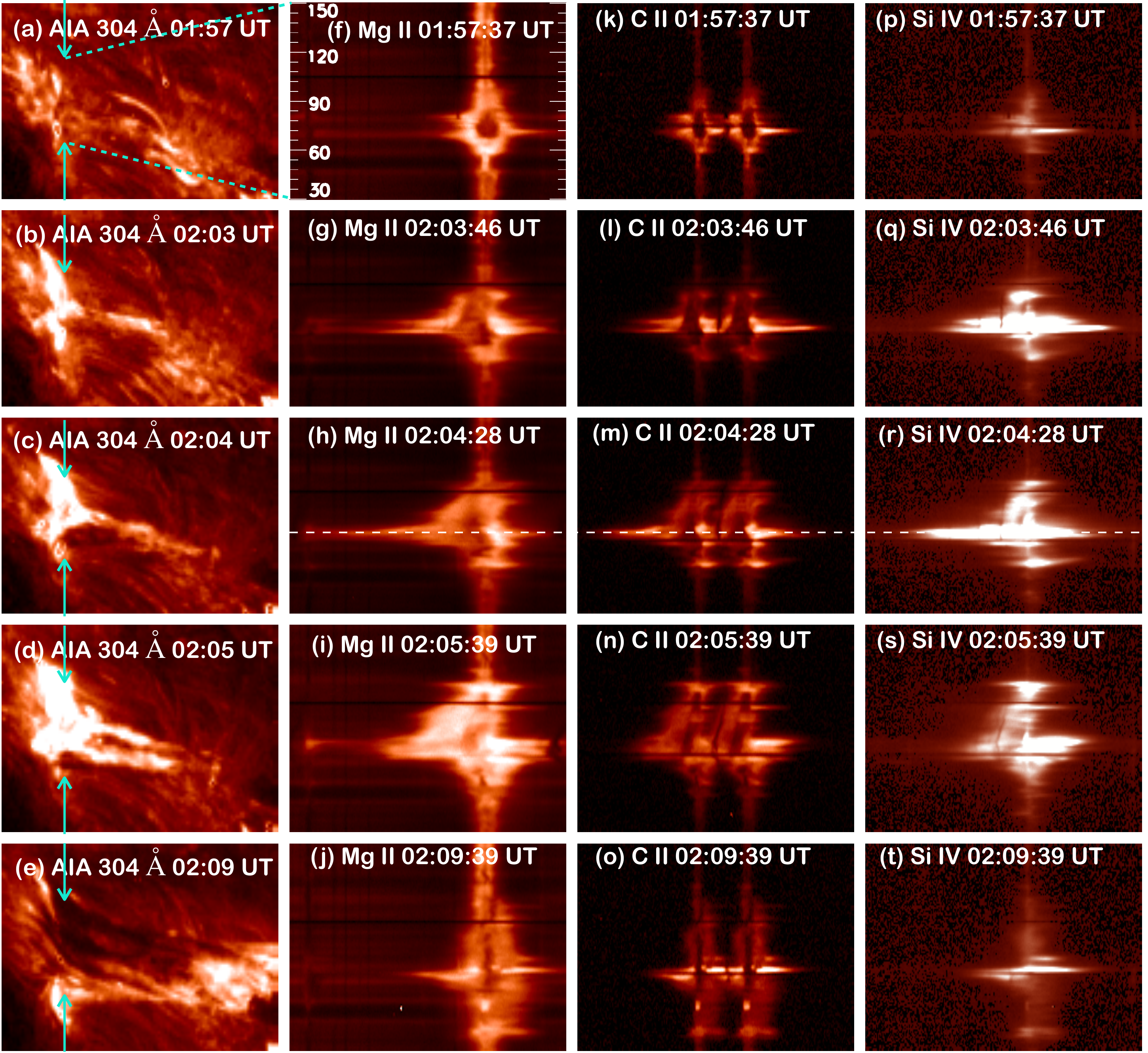}
\caption{Jet reconnection base (UV burst or mini flare) and jet evolution  between 01:57 UT to 02:09 UT (from top to bottom). First column presents the images in AIA 304 \AA.
Second, third, and last columns show IRIS spectra  of the jet reconnection  site (UV burst) at slit 1   
in
the Mg II k 2796.35 \AA\  line , C II doublet, and in Si IV  1393.76 \AA\ respectively. The vertical cyan arrows  and inclined dashed lines (panel (a)) in the first column indicate the position of the slit and the length of the spectra in the three  other columns (30 $< $ y  (pixel) $<$ 150). The white dashed line in panels (h, m, and r) shows the location for the profiles at y=79 pixel, presented in Fig. \ref{threeProfiles}.}
\label{AIA_IRISspectra}
\end{figure*}
Solar jets are commonly observed  with IRIS and the  multi wavelength \emph {Atomospheric Imaging Assembly} \citep[AIA,] [] {Lemen2012} instruments.
The characteristics of such jets vary in large parameter ranges: velocity between 100 to 400 km s$^{-1}$, length  between 50 and 100 Mm \citep{Nistico2009, Joshi2020}.  
IRIS spectroscopic and imaging  observations of jets reveal bidirectional outflows in transition region lines at the base of the jets implying  explosive  magnetic  reconnection processes  \citep{Li2018, Ruan2019}.
Bidirectional outflows in the LOS are detected  by the extended wings in chromospheric and transition line profiles 
\citep{Dere1991,Innes1997,Tian2018,Ruan2019}. While they  are commonly related to  magnetic reconnection,
\citet{Judge2015} proposed a different interpretation based on   Alfv\'enic
turbulence.

It is not clear the relationship between jets, flares, surges, EBs, and IBs. \citet{Young2018} reviewed  separately all these small transient IRIS phenomena with no real physical link between them. It is  useful to be able to provide a scenario where all the pieces of  puzzle (using IRIS line profiles of the different ions) can be integrated in 
a global model of atmosphere for mini flare, jet and IBs. 

The present work is focused on a twisted jet  and at its base, a flare of B6.7 GOES class (that we call mini flare in this study), which occurred on  March  22, 2019 in NOAA  AR 12736 around 02:05 UT. 
The magnetic topology of this region has  been  studied by 
\citet{Joshi2020FR} (from now we will refer this study as Paper I) and is summarized in Sect. \ref{obs}. The present  paper is  mainly focused on the  IRIS data in the frame of AIA 304 \AA\ observations.
Mg II, Si IV, C II spectra and line profiles at the reconnection site of the jet, 
are analysed  leading to   
a sketch of 
dynamical reconnection  (Sect. \ref{Dopplershift}).
In Sect. 
\ref{multi-temps} we discussed  on a possible multi thermal reconnection  model with multi layers from   very deep layers 
in the atmosphere, e.g. at  the minimum temperature region, to the corona. This is how a sandwich model with stratification of multi layers is proposed to explain the observations during the reconnection. In Sect. \ref{dis} we summarized the results and concluded on the multi facets of this mini flare (UV burst and white flare).


 \begin{table*}
\caption{
IRIS observation of NOAA AR 12736 on March 22, 2019.}
\bigskip
\centering
\setlength{\tabcolsep}{10.0pt}
\begin{tabular}{llll}
\hline
\hline
Location& Time (UT)& Raster& SJI\\
\hline
x=709$\arcsec$& 01:43--02:42 &  FOV: ~~~~~~~~~~~~~~~~~~~~ 6$\arcsec$ $\times$ 62$\arcsec$& FOV: 60$\arcsec$ $\times$ 68$\arcsec$\\
 y=228$\arcsec$& &  Step size:~~~~~~~~~~~~~~~~~~~~~
 2$\arcsec$& Bandpass: C II 1330 \AA, \\
 & &  Spatial resolution: 0.${\arcsec}$33& ~~~~~~~~~~~~~~~~~~Mg II 2796 \AA \\
 & & Step cadence: ~~~~~~~3.6 s  & Time cadence: 14 s for each passband \\
\hline
\hline
\label{table1}
 \end{tabular}      
\end{table*}

\section{Instruments and Observations}
\label{obs}
\subsection{
Global evolution of the active region
}\label{obs1}
We report on the observations of a solar twisted jet, a related  surge and a mini flare at the jet base in NOAA AR 12736 located at N09 W60 on March 22, 2019.

Figure \ref{AIA_IRIS} summarizes the observations obtained with the multi-wavelength  filters  of AIA  and   the {\emph Helioseismic and Magnetic
  Imager} \citep[HMI,] [] {Schou2012} aboard \emph{Solar Dynamics Observatory} \citep[SDO,] [] {Pesnell2012}, and IRIS. The jet and the mini flare 
  are observed in all the AIA channels covering a wide range of temperatures (10$^5$ - 10$^7$ K) (animations are attached in AIA 304 \AA\ and 193 \AA).   Figure \ref{AIA_IRIS} (panels a-b) show an example of the mini flare and the jet visible as bright elongated  
  regions in AIA 304 \AA\ (50,000 K) and 193 \AA\ (1.25 MK) filters respectively.
  
  
  The AR  has been  formed by successive emerging fluxes during 24 hours before the jet observations. The AR  magnetic configuration at the time of the mini flare  consists of three     emerging magnetic flux (EMF): an earlier one (orange oval) and  two very active EMFs:  EMF1 (P1-N1) and EMF2 (P2-N2) 
  highlighted  by  the yellow and the red  ovals (Fig. \ref{AIA_IRIS} panel (c), see  also the animation attached for the magnetic evolution with HMI observations). 
  The contours of the longitudinal magnetic field ($\pm$ 300 Gauss) is overlaid  on AIA 304  \AA\ and  
  193 \AA\ images (Fig. \ref{AIA_IRIS} a-b). The  polarity inversion line  (PIL)
 between these two  EMFs  (between P1 and N2 more precisely) 
 is shown by a dashed dark line in panel (a-b).
  The images of second and third rows in Fig. \ref{AIA_IRIS} present a zoom view of the mini flare at the jet base at 02:06:05 UT observed with AIA.  
  In panels (d-f) the jet is seen to develop westwards while the mini flare corresponds to a  North-South arch-shape  brightening along the PIL and 
  a bright point
  in its middle (Fig. \ref{AIA_IRIS} panel f-g).
  
   In the magnetic field evolution (an animation is also attached as MOV3) 
 the negative polarity N2 is sliding along the positive polarity P1 with possibly reconnection  between the two polarities N2-P1 (Fig. \ref{AIA_IRIS} panel i). 
 The green circle indicates the small bipole which is the location of the reconnection site at the jet base corresponding to the `X' point. In the bipole formed by the collision of two polarities belonging to two different magnetic systems, strong shear should exist as it was shown in other cases \citep{Dalmasse2013}.  There is no visible hot loop  in the AIA filters corresponding to  this location.
 After our analysis of the IRIS spectra, we  
  show that the Mg II line  profiles at the jet base  are similar to the  Mg II line profiles in IBs formed in bald patch  region where the magnetic field lines are tangent to the solar surface \citep{Zhao2017}. By analogy it suggests that there is a bald patch (BP) region inside the bipole   (see section 3.4).  It would explain why the bright point  is visible  at the minimum temperature  in the atmosphere e.g in 1700 \AA\ and 1600  \AA\ (Fig. \ref{AIA_IRIS} panel f ). This is in the line of 
  the main conclusion of Paper I, where it has  been demonstrated that the magnetic reconnection initiating the jet started in a BP current sheet which rapidly became an X-null point current sheet.
 
 In the AIA filters (Fig. \ref{AIA_IRIS}
    panels (a-b)) nearly horizontal dark 
  strands forming
   arch filament systems (AFSs) 
    are visible in each side of the mini flare. The EUV emission with wavelengths shorter than the hydrogen 
    Lyman and helium discontinuity continuum (912 \AA, 504 \AA, 228 \AA) 
suffers from the continuum absorption from H I, He I,  and He II due to photoionization  \citep{Heinzel2003,Anzer2005}. Mainly  on the West side overlaying EMF1 and the former EMF (orange oval) 
   the AFS structure 
   has  an  East-West 
   direction.
   It  is common to observe AFS over EMF   during the  emergence of magnetic flux \citep{Schmieder2007}.

\subsection{IRIS observation mode
\label{obs2}}
On March 22, 2019 between  01:43:27 UT and 
02:42:30 UT,  IRIS was targeting the jet base in the NOAA AR 12736 with a field of view of
60$\arcsec$ $\times$ 68$\arcsec$ 
 centered at
x=709$\arcsec$ and y=228$\arcsec$
. When the jet appeared, 
IRIS acquired  slit jaw images (SJIs) in two passbands: SJIs 1330 \AA\ (dominated by the C II lines) 
and  SJI 2796 \AA\ where the emission mainly comes from the Mg II k line. 

Details of the IRIS observations are in Table \ref{table1}.
The co-alignment
between the different IRIS channels was achieved
by using the {\it drot\_map} function of IDL in solar software to correct the differential
rotation. Those SJIs 
 were taken at a cadence of 14 sec for each passband. 
\\


  Simultaneously IRIS performed medium coarse rasters  of 4 steps.
 The raster
step size in x is 
2$\arcsec$ so each spectral raster spans a field of view
of 6$\arcsec$ x 62$\arcsec$ with four positions of the slit. The nominal spatial resolution is 0.$\arcsec$33. 
 During the full observation time it was repeated 250 times.
Calibrated level 2 data are used in this study. Dark current subtraction, flat field
correction, and geometrical correction have been taken into account in
the level 2 data \citep{Pontieu2014}.

IRIS  provides line profiles in  Mg II k  and h lines (2796.4\,\AA{} and 2803.5\,\AA{} respectively),  Si IV (1393.76 \AA, 1402.77 \AA) and C II  (1334.54 \AA, 1335.72 \AA) lines  along the four slit positions (slit length of 202 pixels equivalent to 62 $\arcsec$). The Mg II h and k lines are formed at chromospheric temperatures, {\it e.g.} between 8000 K and 20000 K \citep{Pontieu2014,Heinzel2014b,Alissandrakis2018}. C II is formed around T=30,000 K and Si IV around 80,000 K.
Many other  chromospheric and photospheric lines have been identified in the spectra of the mini flare (see Table \ref{table2}).
 \begin{table*}[ht]
\caption{Identification of  the lines in IRIS wavelength ranges of  C II, Si IV, and Mg II lines observed in the mini flare at the jet base; (bl) means blended.}
\bigskip
\centering
\setlength{\tabcolsep}{14.0pt}
\begin{tabular}{llllllll}
\hline
\hline 
Ion   & $\lambda$ (\AA)    && Ion &$\lambda$ (\AA)&  & Ion &$\lambda$ (\AA)\\
\hline
C II &1334.54       &&  O IV &1399.776&& Mg II triplet   & 2791.6 \\
C II  &1335.72           &&  O IV&1401.163 && Mg II k  &  2796.4  \\
Fe II & 1392.817   &&  Si IV &1402.77  && Mg II triplet & 2797.9\\
Ni II  &1393.33              && O IV &1404.806 (bl) &&   &2798.0 \\
  Si IV &    1393.589                               && Si IV&1404.85 (bl) & &Mg II h &2803.5 \\
 Si IV& 1393.76                            &&S IV &1406.06 &&  & \\
 \hline
 \hline
 \label{table2}
 \end{tabular}
\end{table*}

\subsection{Mini flare observed with AIA and IRIS} 
 \label{miniflare_AIAIRIS}
 An example of 
 IRIS SJIs in 1330 \AA\ and 2796 \AA\, is presented in 
 Fig. \ref{AIA_IRIS} (g-h). West is on the right and East on the left in all the panels with images.
 The FOV of IRIS SJIs includes
  the mini flare (bright point in panels (f-g)) and a part of the wide jet base.  The bright point is considered as 
 as  the reconnection site (or `X' point) at the jet base. The four  positions of the slit scanned the mini flare site around x= 705$\arcsec$ and y = 220$\arcsec$ and the arch-shape  brightening  at the base of the jet (see in panel (h)).
  Globally the structures visible in IRIS SJIs  are similar to those in AIA 304 \AA\ (50,000 K).  The FOV of IRIS has been shifted by 4 arcsec in x axis and 3 arcsec in y axis to be co-aligned with AIA coordinates.  
  
 In AIA 304 \AA\, between  02:04:09 UT to 02:06:09 UT we see that the 
 jet  base has a triangular shape.
 Between the two external sides of the jet's triangular base there are two slightly bright patches in an East-West direction.
In  C II observations (an animation is attached as MOV4) we can follow the formation of small kernels at 02:04:28 UT, 02:05:25 UT,  02:05:39 UT, and 02:06:07 UT, travelling from one side to the other side of the triangle following these bright patches (from east to west). 
  In  AIA 304 \AA\ images,  the development of the surge is well visible in between  02:04 and 02:07 UT (Fig. \ref{AIA_IRIS} panel (e) at 02:06:05 UT). However, the surge is not so well visible in 
  IRIS SJIs 1330 \AA\ and 2796 \AA\  taken at the corresponding times (panels g-h).  This can be explained because the  absorption of the UV emission is only  efficient   for lines with  wavelengths below the hydrogen Lyman continuum limit ($\lambda$ $<$ 912 \AA)  \citep{Schmieder2004}.
  Moreover, the non-visibility of the surge can be due to the large wavelength ranges of the IRIS SJIs filters where  the full line profiles  are integrated and the line emission in the jet was not strong enough. 

    \section{Spectroscopic analysis 
    }
  \label{Dopplershift}


To process the IRIS Mg II h and k data, we used the
spatial and wavelength information in the header of the
IRIS level-2 data and derived the rest wavelengths of
the Mg II k  2796.35 (4) \AA, and  Mg II h 2803.52 (6) \AA\, from the reversal positions of the averaged spectra at the disk.  
For C II and Si IV lines the zero velocity is defined in a similar way  (see Table \ref{table2} for the rest wavelengths used in the present work).
\subsection{IRIS spectra of mini flare}\label{IRIS_miniflare}  
  We show one example of spectroscopic data
  obtained
  between 02:04:28 UT  and 02:04:39 UT with  the four slit positions 1, 2, 3, 4 for the three different elements Mg II, C II, and Si IV (Fig. \ref{spectra}). 
 The correspondence between pixels  along the slit and arcsecs in SJIs is shown as y coordinates of 
  the SJIs (Figure  \ref{spectra} panels e and n).
  The slit  position 1 shown in panel (n) crosses the bright zone between 60 to 105 pixels  (around 210-230 arcsec) corresponding to the jet base.
  In the middle of the zone, the brightest point  
  along the slit   is
  the reconnection site (y $\approx$ pixel 79- 80 corresponding to the position `X' (705$\arcsec$, 220$\arcsec$) in Figs. \ref{AIA_IRIS} (g) and \ref{spectra} (e,n)).
  These spectra  observed around 02:04 UT (panels a-m) correspond to  the onset of the mini flare, 

  
  At the reconnection site the spectra shows very complex structures that we will analyse in the next sections.  We note that in all the slit positions 1-4, similar features are shown, but they are more pronounced in the slit position 1, which seems to be exactly at the reconnection site  for this time. We will mainly restrict our study to the slit position 1.
  It is not really possible to reconstruct  an adequate spectroheliogram image with  only four positions  distant in x  of 2$\arcsec$ each. 
  
\begin{figure*}[ht!]
\includegraphics[width=0.92
\textwidth]{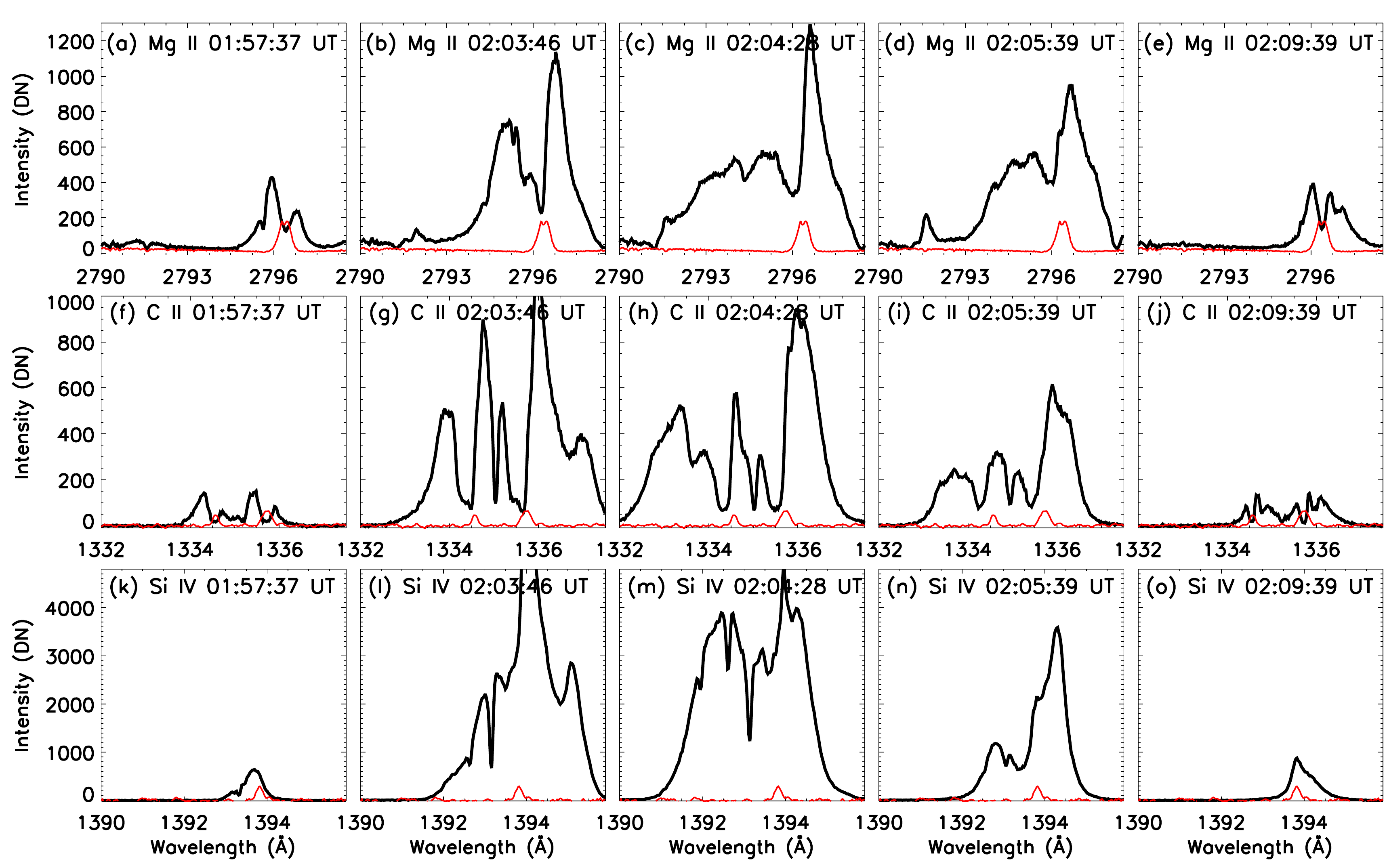}
\caption{{\it (from left to right)} Evolution of jet reconnection site  (UV burst) between 01:57 UT and 02:09 UT through the profiles of the Mg II  k (a to e), C II (f to j), and Si IV  1394  (k to o)  lines observed in slit 1 at y = 79 pixel. The location of the y point in the different spectras is shown in Fig. \ref{AIA_IRISspectra} with a white dashed line.The  considered spectra data are the same as those used for Figure \ref{AIA_IRISspectra}. The small red profile  in each panel is the reference profile. The reference profiles of C II and Si IV  lines are raised by a factor of 5 and 20 respectively.} \label{threeProfiles}
\end{figure*}  
  \subsection{Time evolution of IRIS spectra of mini flare}\label{time_IRIS}  
  We note in Sect. \ref{miniflare_AIAIRIS} that AIA 304 \AA\ images have a better contrast than the IRIS SJIs to show the cool structures visible by absorption.
  Therefore  we co-align carefully the 
  images in AIA 304 \AA\ with the IRIS SJIs in order to indicate
  exactly the position of each pixel of the slit in the AIA 304 \AA\, images  to be able to 
  discuss the evolution of the structures visible in the 304 \AA\, images jointly with the spectra shape of IRIS lines using both coordinates the pixels 
  along 
  the slit  and the AIA  coordinates.
  The evolution of the structures 
 visible  in AIA 304 \AA\ images: mini flare, jet and surge
 are  summarized 
 in five different times in Figure \ref{AIA_IRISspectra}, corresponding to: pre reconnection time (first row), reconnection times (second and third rows), jet base extension (third row), after reconnection time (fifth row) 
     (also see the movie in AIA 304 \AA\ animation attached MOV1). Between the two vertical blue arrows in the AIA images (left column),  a  section of slit at position 1 is located. The right columns present the spectra in this section for the three  elements Mg II, C II and Si IV.
   Table \ref{table4} gives a  detail about the characteristics of these  typical profiles in the four slit positions during the different phases of the jet time observations.  They are changing very fast and it is rather complicated to analyse all of them. We will nevertheless focus  on the profiles which  are important for understanding the reconnection process in Sects. \ref{comparison} to \ref{sec:tilt}
   and be able to proceed to a dynamical model of reconnection (Sect. \ref{sketch}).
   
  \subsection{Characteristics of the spectra: spatio-temporal analysis of IRIS spectra}
  \label{comparison}
  For each  IRIS spectra shown in Fig. \ref{AIA_IRISspectra} we select the pixel 79  corresponding to the reconnection site and  draw the   line profile  of the three elements  for the five times (see Fig. \ref{threeProfiles}). The line profiles help  to  interpret  the nature and evolution of the structures during the different phases of the reconnection. We focus our analysis of Mg II and C II chromospheric lines in the following subsections during the three phases of the jet reconnection. Then we analyse in details the line profiles of the three elements  during the reconnection times (Sect. \ref{UV_burst}).
  \subsubsection{Pre reconnection time}\label{pre}
 Around 01:57 UT in AIA 304 \AA\ image, tiny vertical bright 
 areas along the inversion line are visible (see panel (a1) in Fig. \ref{AIA_IRISspectra}). The corresponding  C II and Mg II  spectra show  very large central dip which could represent the presence of cool material at rest which  absorbs the incident radiation (Fig. \ref{AIA_IRISspectra} panels (b1 and c1)) and the corresponding line profiles in  
Fig. \ref{threeProfiles}  panels (a and f).

\begin{figure*}
\centering
\includegraphics[width=1.0 \textwidth]{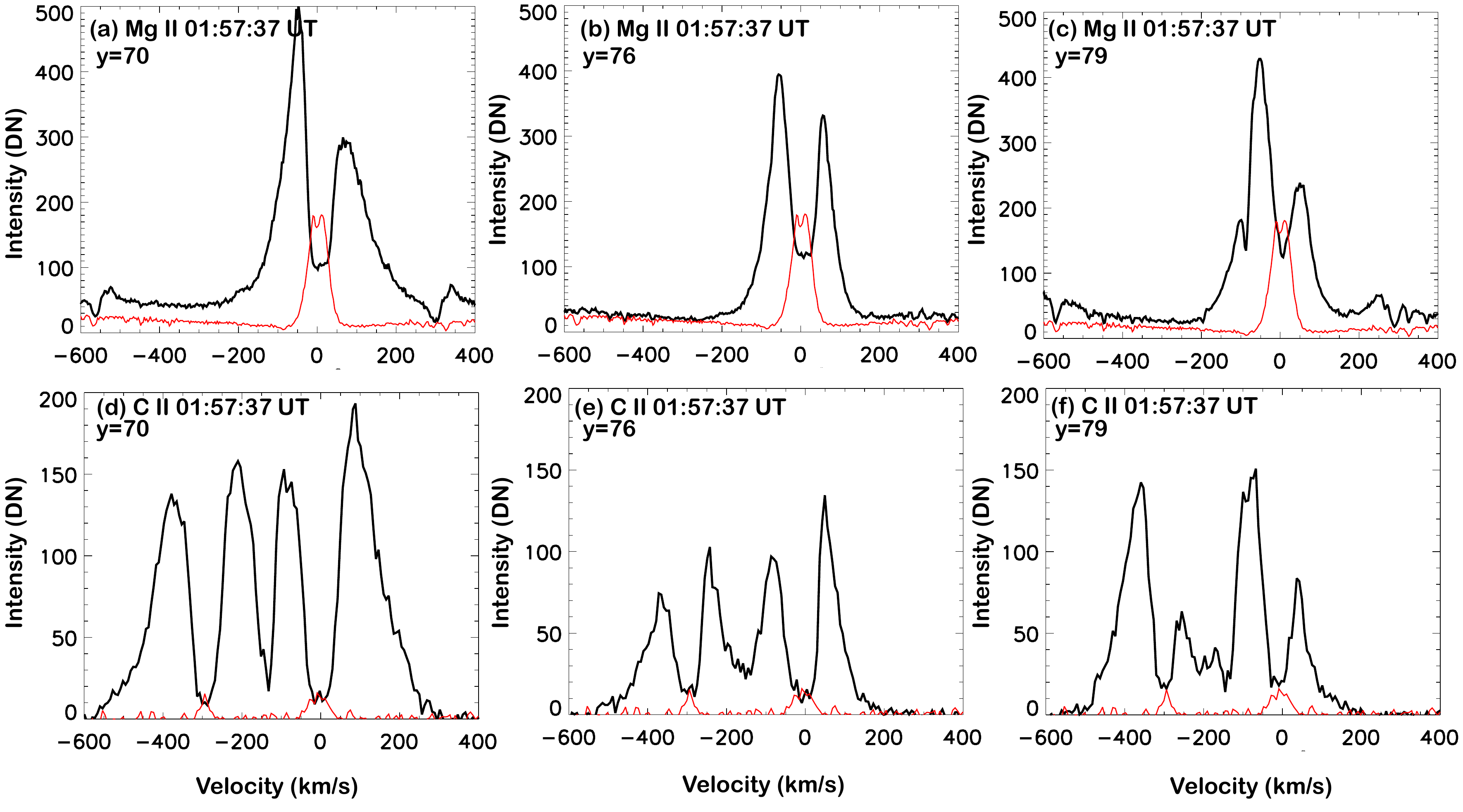}
\caption{Comparison of the profiles of Mg II k (top row) and  the two C II (bottom row) lines in
three pixels along the slit at the position of the UV burst 
at 01:57:37 UT before the burst (Fig. \ref{AIA_IRISspectra} f). 
The x axis unit is velocity to compare the shift of the peaks of the Mg II and C II lines. 
We note the same   large distance between the two peaks of the Mg II and C II lines, which could be due to the presence of cool material in the region of the reconnection before the reconnection.  The red profiles are reference profiles, which were used to determine the rest wavelengths.}
\label{Mg_CII_lines}
\end{figure*}

The Mg II k and C II line profiles at this  position  and around (pixels= 70, 76, 79) are presented  again  but with a zoom  and   
with the x-axis in Dopplershift units in km s$^{-1}$  (Fig. \ref{Mg_CII_lines}). They are  very broad  with a central dip (FWHM more than 1 \AA\  which corresponds to +/-50 km/s) while the peaks of  the Mg II and C II lines are equally distant (100 km s$^{-1}$). The central dip    would imply that cool material absorbed the incident radiation 
more or less at the rest. Such cool material could be due to parts of  arch filaments  
trapped in  the magnetic field lines  between the two EMFs (EMF1 and EMF2) in the vicinity of the bright point region before the reconnection.

 \subsubsection{During the reconnection time}\label{during}
 Around 02:03 - 02:05 UT the  mini flare (UV burst at the `X' point) starts in the middle of this bright area  with  the onset of the jet  ejection (Fig. \ref{AIA_IRISspectra} with  304 \AA\,images in panels (b,c,d), Mg II spectra in panels (g,h,i) and C II spectra  in panels (l,m,n)) and their corresponding line profiles in 
Fig. \ref{threeProfiles}  panels (b,c,d and g,h,i). The bright jet is obscured by a surge, a  set of dark (cool) materials in front of it; both the jet and the surge are extending toward the West at the same time. In AIA 304 \AA\  image at 02:05 UT the jet is extended along two bright  branches with  a dark area in between. During this time the spectra show very broad blue  wings along the slit in the same zone $\pm$ 10 arcsec around y=220$\arcsec$. 
 At y= 79 pixel (approximately at 220$\arcsec$), the profiles in all the lines are the  most extended (Fig. \ref{threeProfiles}  panels (b,c,d and g,h,i)). We notice that the spectra along the slit show a tilt at the northern bright branch (Fig. \ref{AIA_IRISspectra} panels i,n), which could indicate some rotational motion there (see Sect. \ref{sec:tilt} for more details).
 The wavelength positions of the dark absorption core of the Mg II and C II line profiles 
 along the slit show a clear zigzag pattern of the blue and red shifts  which could correspond to cool plasma motion with different  velocities  along the slit.  
  In the next sections (Sects. \ref{UV_burst} and \ref{sec:cloud}) we analyse the profiles of the three lines
  to obtain  quantitative values of the Dopplershifts of the plasma in  the  reconnection zone of the jet.
 \subsubsection{After reconnection time}\label{after}
 At 02:09 UT, long dark East-West filament structures in the North of the reconnection site are observed
 in the 304 \AA\ images (see Fig. \ref{AIA_IRISspectra} (e)).
 Their corresponding spectra show a dark core and  weak emission in the   red  wings  all along the slit (y(pixel)=70-105) which could correspond to  cool plasma absorbing the red peak emission of the jet. This cool plasma may be plasma of the surge or to the arch filament system  going  away of the observer with Dopplershifts of less than 30 km s$^{-1}$ (Fig. \ref{AIA_IRISspectra} (j and o)). 
 We note an enhancement of the continuum emission close to the Mg II lines which could correspond to enhancement of the Balmer continuum like in white light flares (Fig. \ref{AIA_IRISspectra} panel i) \citep{Heinzel2014}.
  
  \subsection{UV burst in `X' point- large blueshift}
  \label{UV_burst}

The IRIS slit at position 1 
crosses the mini flare (UV burst at `X' point) so that the
spectra along the slit bring many information about the dynamics of the UV burst as explained in the previous section (Sect. \ref{comparison}) and in  
Fig. \ref{AIA_IRISspectra}. Figure \ref{threeProfiles}
shows the evolution of the UV burst 
  using the three lines (Mg II, C II, and Si IV) profiles for  y(pixel)=79 (220$\arcsec$) with a time scale of  one  minute.  
 The  profiles change very fast on this time scale.  We  analyse  these profiles at each time in order to derive  the characteristics   (velocity and temperature) of  structures which are  integrated along the LOS.

At 02:03:46 UT  in very localized pixels  inside the burst   Mg II, C II and Si IV  profiles have   more or less symmetrical profiles with high peaks  with extended blue and red wings ($\pm$ 200 km s$^{-1}$)  (Fig. \ref{threeProfiles} panels 
b, g, i and in Annex: Fig. \ref{burst_MgII} a and \ref{burst_SiIV}). 
With  such  extended wing  profiles in a few pixels we may think of bilateral outflows  of reconnection 
\citep{Ruan2019}. Such  outflows 
with   super Alfv\'enic speeds were  observed in a    direction perpendicular to the jet initiated by the reconnection like in our observations. 

\begin{figure*}
\centering
\includegraphics[width=1.0 \textwidth]{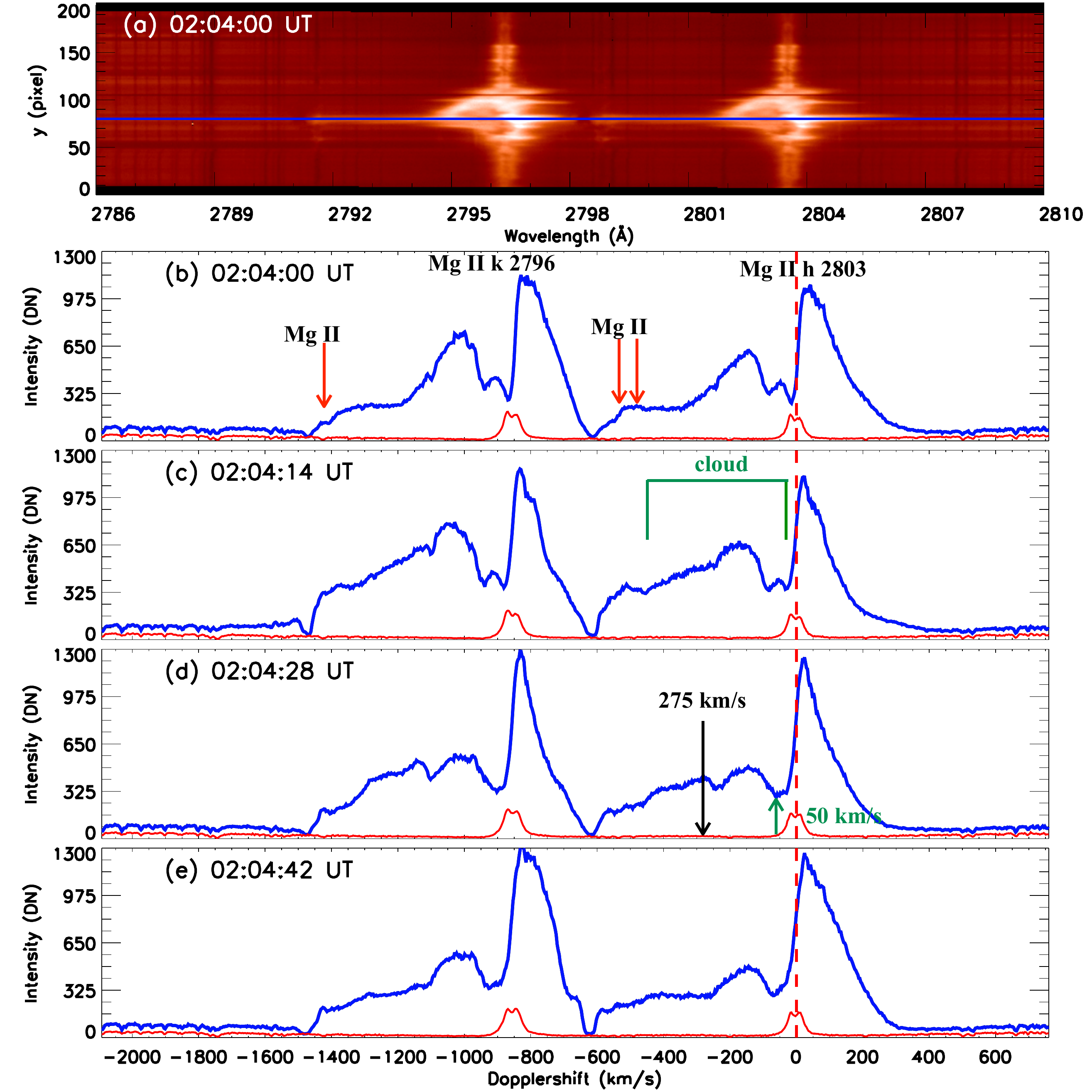}
\caption{(a) Mg II spectra at 02:04:00 UT along the slit position 1 before the UV burst  (Fig. \ref{AIA_IRISspectra} g). Panels (b-e) from top to bottom:  evolution of the Mg II k and h line profiles (for  y = 79 pixel in  panel (a))   showed by a profile every 14 sec during less than one minute time (02:04:00-02:04:57 UT) (Fig. \ref{AIA_IRISspectra} h). The emission of the Mg II triplet line profiles--one at 2791.6  \AA\ and other two at 2799 \AA\, are indicated by red arrows in panel (b). 
The horizontal green brace points  the low intensity value of the Mg II blue peak, signature of the possible  presence of an absorbing (red arrow in panel d) and emitting radiation cloud (black arrow in panel d)  along the line-of-sight.
The reference profiles are shown in red. The vertical red dashed lines in panels (b-e) show the position of rest wavelength at 2803.6 \AA. The wavelength range in panel (a) and (b-e) is same in wavelength unit for panel (a) and the corresponding Dopplershift unit in panel (e).}
\label{MgII_evolution}
\end{figure*}

\begin{figure*}
\centering
\includegraphics[width=1.0 \textwidth]{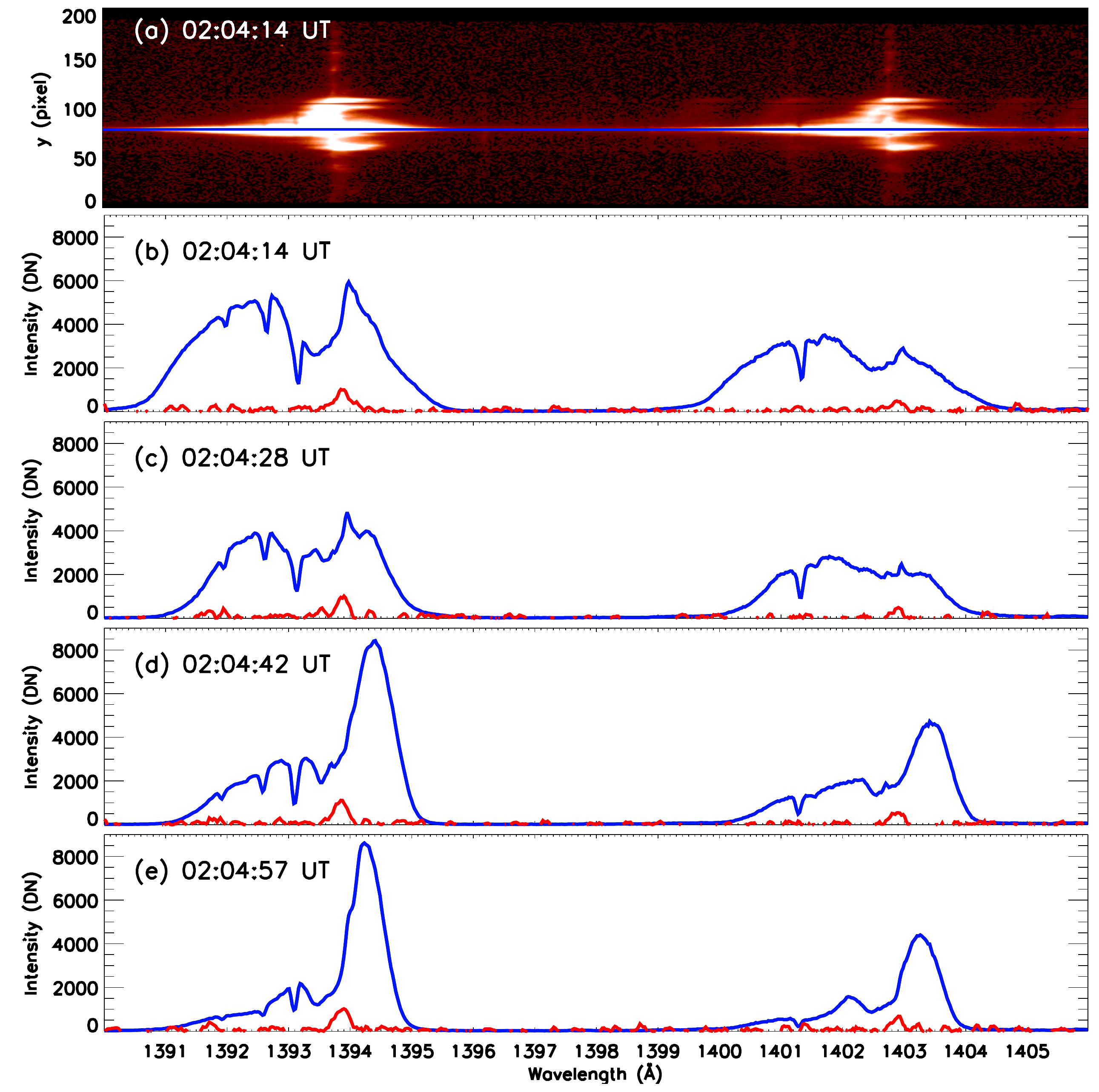}
\caption{(a) Si IV  spectra at 02:04:14 UT 
at the start of  the UV burst (Fig. \ref{AIA_IRISspectra} a). Panels (b-e): from top to bottom: fast evolution  during less than one minute,   
one  Si IV  profile every 14 s  at slit position 1 between  02:04:14 UT and 02:04:57 UT. The profiles from panel (b-e) are taken at y(pixel) = 79, shown as blue horizontal line in panel (a). The reference profiles are shown in red. 
.}
\label{SiIV_evolution}
\end{figure*}

For the time of reconnection  around  02:04:28 UT, Mg II, Si IV and C II line spectra are presented for the four slit positions (Fig. \ref{threeProfiles}).
 The profiles at this time at y=79 pixel exhibit a  high  peak of   
  emission with strong blue shift extended wing (see third column (panels c,h,m) in Fig. \ref{threeProfiles}) although 
  the evolution of the profiles are shown with a low cadence. In Figures \ref{MgII_evolution} and \ref{SiIV_evolution} we show  the details of the fast evolution of the UV burst between 02:04:14 UT and 02:04:28 UT taking advantage of 
  the high cadence of IRIS (14 sec).  
  
  The Mg II profiles of the UV burst during  this time scale did not evolve drastically, contrary to the Si IV profiles in the same time interval. The Si IV profiles are very broad during the UV burst maximum with a FWHM of the order of 4 \AA\ (Fig…\ref{SiIV_evolution}.  A few seconds later at  02:04:57 UT the Si IV profiles consist only  of  one peak  with a FWHM of 1 \AA\, and  an  intensity increasing about a factor of 100. 
  
  We analyse  first the Mg II profiles of the  mini flare  (UV burst)  to understand the composition  and the dynamics of the plasma along the LOS.
  The zero velocity is  defined, as we explained earlier, by  the dip in the reverse profile of Mg II line profile observed in the chromosphere.
  The Mg II profiles are  also very broad with a FWHM  of the order of 5 \AA\, and asymmetric with a high red peak and a very extended blue wing (Fig. \ref{MgII_evolution}).  
 The  blue peak  is much lowered compared to the red one. This characteristic of the profiles can be produced by the  absorption of the blue peak  emission by a cloud  centered around 50 km s$^{-1}$. In the far blue wing an emission  is detected until -5 \AA, which might  come from a  second  cloud centered around a higher value (intuitively determined around - 275 km s$^{-1}$) and  the emission of 
 the Mg II triplet at 
 2797.9 and 2798 \AA\ which are  effectively at -5 \AA\ from Mg II h. All the  Mg II triplet lines have been identified in the spectra (Table 2).
  
\begin{figure*}
\centering
\includegraphics[width=1.0 \textwidth]{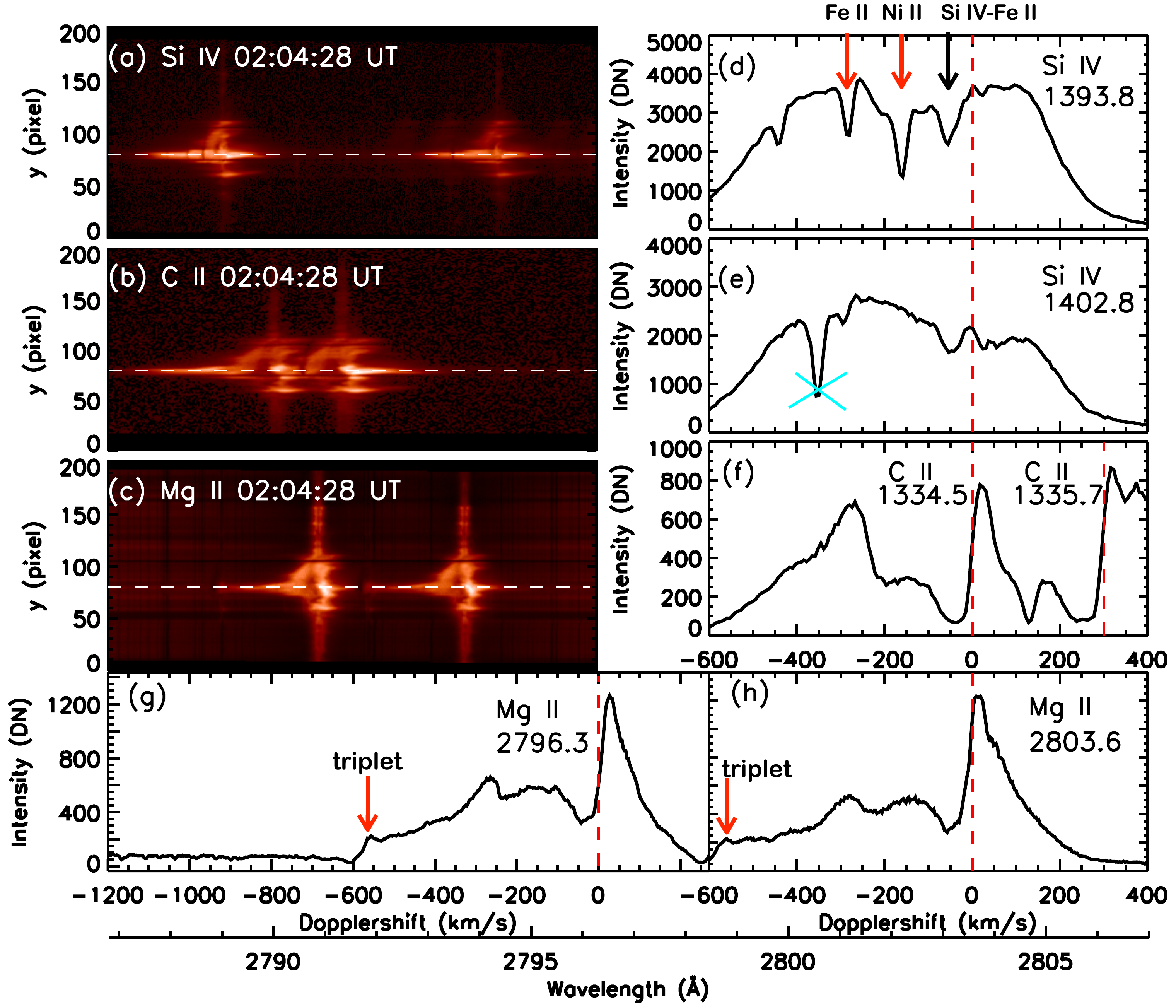}
\caption{Spectra of the jet base (UV burst) showing  the extended blue wing of 
 Si IV  line (panel (a)), C II line  (panel (b)) and Mg II  line (panel (c)) at 02:04:28 UT (Fig. \ref{spectra} left column), the white horizontal dashed lines in these three panels indicate the position where the profiles are drawn in panels (d-h). In panels (d-h), the horizontal wavelength axis is shown as Dopplershift relative to the rest wavelength 
  in each window. The Doppler-shift scale in
all panels is the same for the comparison of the Dopplershifts in three element lines.
 We note the presence of dips in the Si IV profiles corresponding to absorption of chromospheric lines (Ni II 1393.33 \AA\, and Fe II 1393.589 \AA) indicated by red arrows  in (d). The black arrow in panel (d) indicates  a dip due to self-absorption of Si IV line blended by a Fe II line. The cyan cross in panel (e)  points a dip with missing data in the spectra. The dashed vertical red lines indicate the zero velocity. The red arrows in panels (g-h) indicate the emission of Mg II triplet lines.} \label{MgII_SiIV}
\end{figure*}

  The profiles of   Mg II, C II, Si IV lines at the UV burst 
  plotted in Dopplershift units relative to the rest
wavelength 
  show similar velocities, 
  which indicates that they correspond to real plasma moving with high flow speed (see Fig. \ref{MgII_SiIV}).
  Although the extended  blue wings could also  be  interpreted as due to a gradient of velocity inside the cloud along the LOS instead of a moving cloud. In the next section we apply a cloud model technique to have quantitative results.
  
  \subsection{Cloud Model Method for Mg II lines}
  \label{sec:cloud}

 Cloud model method was first introduced by \citet{Beckers1964} for understanding asymmetric line profiles in the chromosphere. The structure overlying the chromosphere is defined by four constant parameters: optical thickness, source function, Doppler width  and radial velocity.  Moreover, \citet{Mein1988} developed the  cloud method by  considering non constant  source function and velocity gradients. Therefore this technique was  applied for different structures  with large velocities, mainly observed in the H$\alpha$ line, {\it e.g.} post flare loops \citep{Gu1992,Heinzel1992},  spicules on the disk \citep{Heinzel1994}, and  atmospheric structures in the quiet-Sun \citep{Mein1996,Chae2020}, even using   
  multi clouds (\citet{Gu1996,Dun2000},  see review by \citet{Tziotziou2007}). 
  This new development  allows to derive dynamical models in the chromosphere  \citep{Heinzel1999}. This technique is valid for a  chromospheric structure with a large discontinuity (e.g. a high  radial velocity), overlying the chromosphere along the LOS.

  Recently \citet{Tei2018} 
 applied the cloud model  technique with constant source functions to  the Mg II lines  which present complex profiles because of their  central reversal. Considering  multi clouds, Mg II complex profiles of off-limb spicules were  successfully fitted \citep{Tei2020}.
 
 Cloud model technique applied to Mg II lines allows us to  unveil the existence of moving clouds over the chromosphere. For the present analysis, this is how during the peak phase of the reconnection two clouds
  overlying the region of reconnection are considered to fit the asymmetric Mg II profiles observed in the UV burst region.
  Mg II  asymmetric line profiles are assumed to be the result of the presence of two overlapping clouds 
{\it c}1 and {\it c}2 located above a background atmosphere along the LOS. We suppose the background atmosphere is symmetric with high peaks in the Mg II lines.
We consider a situation where the cloud {\it c}2 is located above the cloud {\it c}1 along the LOS.
Assumptions relative to the two clouds are as follow;
 \begin{enumerate} 
 \item {The absorption profile of a cloud  has a Gaussian shape.}
 \item {The  two clouds have generally different physical properties.} 
 \item {The source function, 
 the LOS velocity, the temperature, the turbulent velocity in each cloud are independent of depth (constant in the cloud).}
  \end{enumerate} 
  
  The first assumption concerning the absorption profile shape  described by a Gaussian function is defended by the following argument. We are mainly interested in the Doppler-shifted feature and we derive its velocity using the standard cloud model.
  In the center of this Doppler feature, the line absorption profile is very well described by the Gaussian function (the Voigt function is approximately Gaussian in the line core and Lorentzian in the wings) and thus we use it for simplicity.\\
\begin{figure}
\centering
\includegraphics[width=0.45 \textwidth]{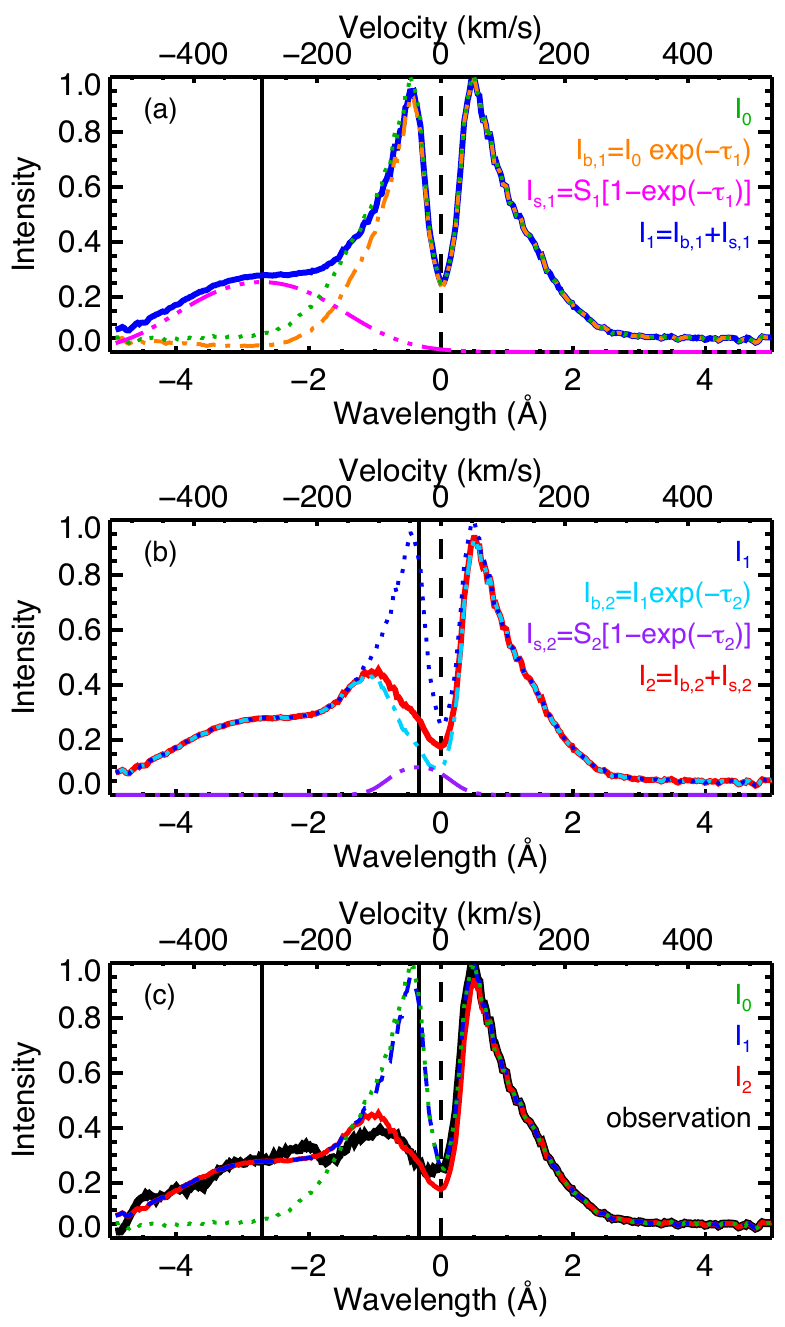}
\caption{
Two-cloud model calculation of the Mg II k 
line profile at 02:04:28 UT for  pixel y=79 (Fig. \ref{MgII_evolution} d). We consider a situation where the cloud {\it c}2 is located above the cloud {\it c}1 along the LOS.
(a) Detail of the $I_1$ profile.
Dotted (green) line shows $I_0$, the background intensity profile (symmetric) made from the red-side of the observed intensity profile (part of the mini flare symmetrical profile).
Dot-dashed (orange) line shows $I_{b, 1}\equiv I_0{\rm e}^{-\tau_1}$, the background intensity attenuated by the cloud {\it c}1.
Dot-dot-dot-dashed (pink) line shows $I_{s,1}\equiv S_1 [1-{\rm e}^{-\tau_1}]$, the emission from the cloud {\it c}1.
Solid (blue) line shows $I_1=I_{b,1}+I_{s,1}$, the resulting intensity profile when there are only the background atmosphere and the cloud {\it c}1 along the LOS.
(b) Detail of the $I_2$ profile.
Dotted (blue) line shows $I_1$, the intensity from the background and the cloud {\it c}1 (as "background" intensity profile for the cloud {\it c}2).
Dot-dashed (light blue) line shows $I_{b, 2}\equiv I_1{\rm e}^{-\tau_2}$, the intensity from the background and the cloud {\it c}1 attenuated by the cloud {\it c}2.
Dot-dot-dot-dashed (purple) line shows $I_{s, 2}\equiv S_2[1-{\rm e}^{-\tau_2}]$, the emission from the cloud {\it c}2.
Solid (red) line shows $I_2=I_{b, 2}+I_{s, 2}$, the resulting (modeled) intensity profile when there are the background atmosphere, the cloud {\it c}1, and the cloud {\it c}2.
(c) Comparison of observed (solid, black) and modeled (solid, red) profiles.
The background profile $I_0$ (dotted, green) and the profile $I_1$ (dashed, blue) are also shown for comparison. The vertical solid black lines in each panel show the LOS velocity  values of the considered clouds.
}
\label{cloud1}
\end{figure}
The total observed intensity $I_m(\Delta\lambda)$ emitted, when there is one cloud (m=1) or there are two clouds (m=2) on the background atmosphere of intensity $I_0(\Delta\lambda)$ along the LOS, is given by the relation
\begin{equation}
I_m(\Delta\lambda)=I_{m-1}(\Delta\lambda){\rm e}^{-\tau_m(\Delta\lambda)}+S_m[1-{\rm e}^{-\tau_m(\Delta\lambda)}],
\label{eq:m-cloud}
\end{equation}
where
$S_m$ is constant the source function  and
\begin{equation}
\tau_m(\Delta\lambda)\equiv \tau_{0, m}\exp\left[-\left(\frac{\Delta\lambda-\Delta\lambda_{{\rm LOS}, m}}{\Delta\lambda_{{\rm D}, m}}\right)^2\right]
\end{equation}
 is the optical thickness of the cloud {\it c}1 (m=1 case) or {\it c}2 (m=2 case) with the Doppler width
\begin{equation}
\Delta\lambda_{D, m}\equiv \frac{\lambda_0}{c}\sqrt{\frac{2k_{\rm B}T_m}{m_{\rm Mg}}+V_{{\rm turb}, m}^2}.
\end{equation}
Here,
$\Delta\lambda =\lambda-\lambda_0$ is the difference between the wavelength, $\lambda$, and the rest wavelength of the Mg II 
line considered, 
$\Delta\lambda_{{\rm LOS}, m}\equiv\lambda_0 V_{{\rm LOS}, m}/c$ is the shift of wavelength corresponding to the LOS velocity of the cloud of number {\it m}, $V_{{\rm LOS}, m}$ ($c$ is the light speed);
$T_m$ and $V_{{\rm turb}, m}$ are the temperature and the turbulent velocity of the cloud of number {\it m}, respectively;
$k_{\rm B}$ is the Boltzmann constant;
$m_{\rm Mg}$ is the atomic mass of magnesium.
Combining the equation (\ref{eq:m-cloud}) of $m=1$ and the one of $m=2$, the total observed 
intensity $I_2(\Delta\lambda)$ emitted by two clouds is given by the relation
\begin{equation}
\begin{aligned}
I_2(\Delta\lambda)=I_0(\Delta\lambda){\rm e}^{-\tau_1(\Delta\lambda)}{\rm e}^{-\tau_2(\Delta\lambda)}+S_1[1-{\rm e}^{-\tau_1(\Delta\lambda)}]{\rm e}^{-\tau_2(\Delta\lambda)}\\
+ S_2[1-{\rm e}^{-\tau_2(\Delta\lambda)}].
 \end{aligned}
\end{equation}
In the present work, we adopt $T_1 = T_2 = 10^4$ K since the cloud temperatures do not affect the result as long as we use a temperature lower than 20000 K  at which Mg II is ionized.
This is because Mg atom is relatively heavy and the thermal width is small compared to the non-thermal velocity in this situation.
For the background intensity, $I_0(\Delta\lambda)$, we use a symmetric line profile constructed from the red-side of the observed  profile, as done by  \cite{Tei2018}.
In addition, $\alpha_m$ is defined as the ratio of the source function of the cloud of number {\it m} to the background intensity at the line center [$\alpha_m\equiv S_m/I_0(\Delta\lambda=0)$].
We consider a situation where a low velocity component is in the foreground ({\it c}2) along the LOS in order to lower the  peak intensity as it is observed.
Figure \ref{cloud1} shows the result of a two--cloud model fitting. The values of the free parameters are summarized in Table \ref{tab:cloud}.
Two clouds have been detected, one with strong blueshifts (-290 km s$^{-1}$) and the other with a large optical thickness but lower blueshift (-36 km s$^{-1}$); these values are not far from our approximate estimation (Sect. \ref{UV_burst}). Note that here the radiative transfer is completely treated under the above conditions. The turbulent velocity derived for the cloud {\it c}1 is large (150 km s$^{-1}$). This could correspond to the existence of a large velocity gradient inside the cloud, which has not be considered in the assumptions where on the contrary all the parameters are constant. On the other hand,  we may note  that the assumption of a  symmetrical  Mg II profile background for  the flare  does not influence  the fast cloud existence, since the wavelength range of this component is very far in the wing (see Fig. \ref{cloud1}). 

  \begin{table}[ht]
\begin{center}
\caption{Results of two-cloud modeling (c1 and c2)}
\begin{tabular}{|c|c|c|c|}
\multicolumn{4}{c}{Cloud {\it c}1}\\
\hline
$\alpha_1$ & $\tau_{0, 1}$ & $V_{{\rm LOS}, 1}$ & $V_{{\rm turb}, 1}$ \\ \hline
1.6 & 0.99 & $-$290 km s$^{-1}$ & 150 km s$^{-1}$ \\ \hline
\multicolumn{4}{c}{Cloud {\it c}2}\\
\hline
$\alpha_2$ & $\tau_{0, 2}$ & $V_{{\rm LOS}, 2}$ & $V_{{\rm turb}, 2}$ \\ \hline
0.5 & 1.6 & $-$36 km s$^{-1}$ & 50 km s$^{-1}$ \\ \hline
\end{tabular}
\label{tab:cloud}
\end{center}
\end{table}


\begin{figure*}[ht!]
\centering
\includegraphics[width=1.0 \textwidth]{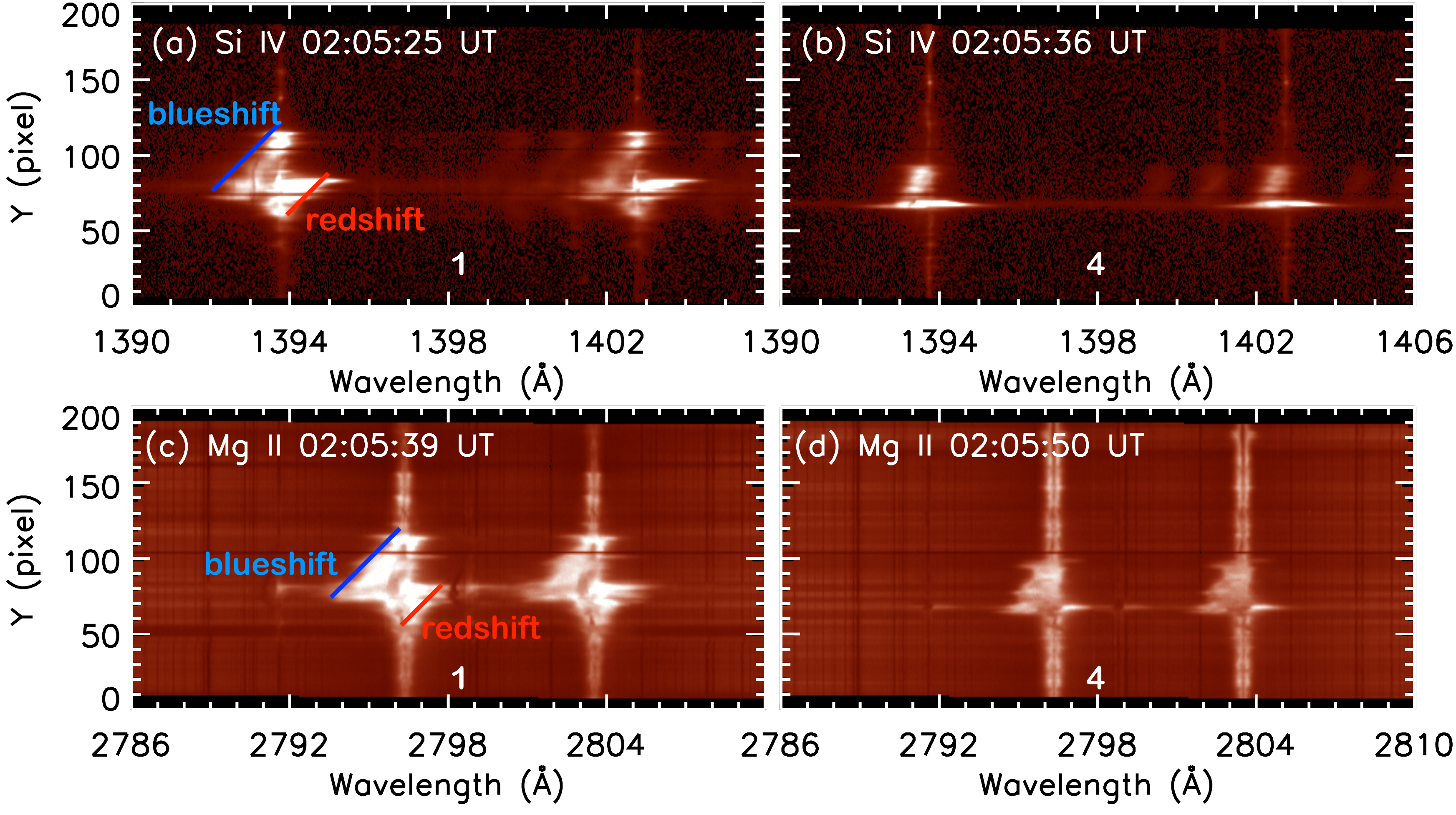}
\caption{Tilt observed in the Si IV and Mg II spectra during the GOES flare time at slit positions 1 and 4,  distant of 6 arcsec (panels a-b for Si IV lines, panels c-d for Mg II lines) (see Fig. \ref{AIA_IRISspectra} panels d, i, s).
The blue and redshifts are shown with the solid lines in the spectra of Si and Mg at slit position 1.
}\label{gradient_spectra_Mg}
\end{figure*}

\begin{figure*}
\centering
\hspace{3cm}
\includegraphics[width=1.0 \textwidth]{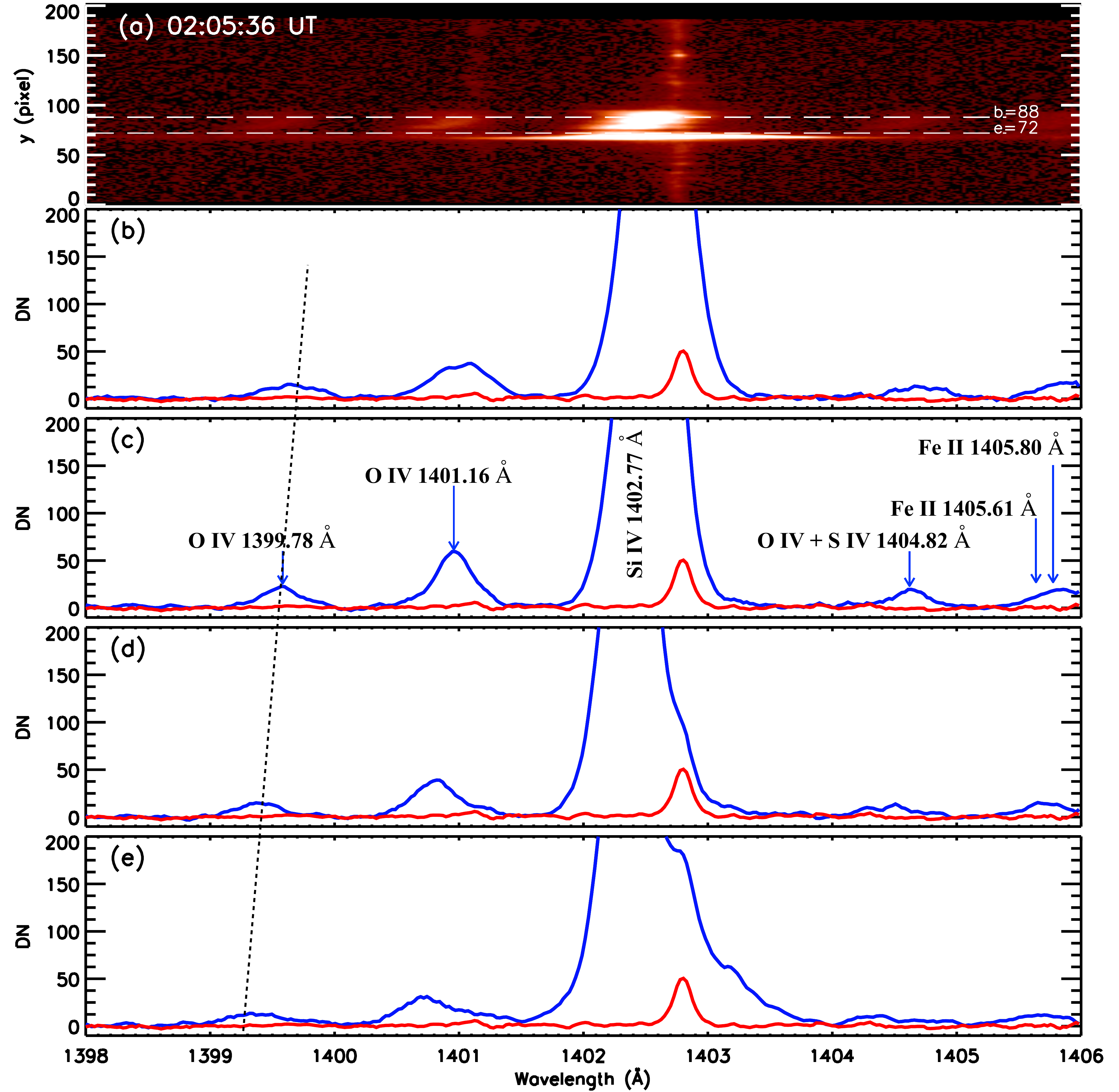}
\caption{Si IV profiles in slit  position 4 at 02:05:36 UT 
for pixels  y = 72 (panel e), y = 76 (panel d),
y = 82 (panel c), and 
y = 88 (panel b). The positions of these pixels lie between the two dashed lines in panel (a).  The vertical dashed line in panels (b-e) represents the tilt of the spectra. In position 4 the profiles are mainly blueshifted above a pixel with bilateral flows. The nominal wavelengths of each identified  line are written on the top of the lines and are listed in  Table \ref{table2}. Reference profile is drawn in each panel in red.} 
\label{SiIV_profiles_slit4}
\end{figure*}

%
%

  \subsection{Spectral tilt profiles}
  \label{sec:tilt}
  Spectra 
  of the Mg II, C II, and Si IV lines show a spectral
tilt at 02:05:39 UT (Fig. \ref{AIA_IRISspectra} panels (i,n,s)). Figure \ref{gradient_spectra_Mg} details the spectra of Mg II and Si IV lines for different times 02:05:25 UT and 02:05:39 UT  in two slit positions distant of 6 arcsec. The tilt is visible in these two positions, the profiles  have dominant
red wings in the southern part of the brightening
(y(pixel) = 50 to 79), they become roughly symmetric
in the middle of the brightening (y = 79) with large extended blue wing nevertheless
and show dominant blue wings in the northern
part  with decreasing blueshifts  until being symmetrical profiles (y = 79 to 120). 
 The tilt is well visible in  O IV lines, lines in emission  identified in the vicinity of Si IV 1402.77 \AA\, (see Table \ref{table2} and Fig. \ref{SiIV_profiles_slit4}).
We could quantify the displacement of the line according to the position along the slit.


These types of spectra
are well known and are typically associated
with twist \citep{Pontieu2014} or rotation \citep{Rompolt1975,Curdt2012} or the presence of 
 plasma in helical structures \citep{Li2014}. 
The tilt  observed in our  spectra can be explained by  the presence of an helical structure at the base during the reconnection process
due to  transfer of twist from a flux rope in the vicinity of the jet, to the jet (Paper I).

  \subsection{Sketch of the dynamical plasma}
  \label{sec:sketch}
  We propose  a sketch to explain the dynamics of the plasma during the reconnection (Fig. \ref{sketch}).  We draw the field lines in solid black lines, the flow directions are indicated with  blue or red  arrows, with solid thick arrows for blueshifts or redshifts. Before the  jet onset (panel a) the magnetic topology of the region consists of  two emerging flux:  EMF1 and EMF2  overlaid  by AFS (see Sect. \ref{obs1}). Between EMF1 and EMF2 the bipole (P1-N2) is located, where the reconnection takes place.  Just before the  reconnection (panel b), there is the formation of a suspected bald patch region in the middle of the bipole with cool plasma trapped inside (see Sect. \ref{time_IRIS}).
 At the time of reconnection (panel c) cool plasma clouds are ejected  with strong blueshifts (see Sect. \ref{comparison}). As the region  is located  at W 60, the clouds with blueshifts  are in fact  ejected over the emerging flux EMF2.  At the same time the bald patch is transformed in an `X' null-point current-sheet with bilateral outflows (panel c). Simultaneously there is the  ejections of the jet and  surge  on the right side of the reconnection site over EMF1 
   (panel d). After the reconnection long AFS and hot loops are formed  overlying the region (panel e).
  This sketch is in the line of the cartoon proposed in Paper I for explaining the reconnection in a X-current sheet.

 \section{Multi temperature layers}
 \label{multi-temps}
  \subsection{Presence of cool material over hot atmosphere}
   \label{cool_over}
During the reconnection time (around 02:04 UT) the IRIS spectra in the mini flare show the presence of cool material over hot plasma like in IBs (Sect. \ref{comparison}).
 This is how the Mg II  large extended blueshift profiles have been  interpreted by the existence of two cool clouds over the reconnection site at the time of the reconnection.  One part  of the trapped cool material could be  ejected with   a low velocity  while the other part   is ejected with  a fast  upward velocity during approximately one minute.
  Cool material is  propelled  to a distance  of 20,000 km along the LOS in one minute (300  km s$^{-1}$ $\times$ 60 sec).  Moreover the large dark dip in the Mg II line centers at this time   could be considered as the cool plasma of the  surge which exhibits low Dopplershifts but high transverse velocity (see also the animation attached as MOV1 in AIA 304 \AA).

  The presence  of such cool plasma over the heated atmosphere at the reconnection site is also  confirmed by the presence of chromospheric lines  striping the Si IV profiles. 
   Si IV  1393.8 \AA\, profiles in the UV burst  reveal 
  that 
  the presence of absorption
lines from singly ionized species (Fe II and Ni II)
(Figs. \ref{MgII_SiIV}d, \ref{burst_SiIV}e and Table  \ref{table2}).  
The
presence of  such lines superimposed on emission lines implies that cool chromospheric  material
is stacked on top of hot material.
The cool material would  come from the bald patch region when the  magnetic field lines were tangent to the solar surface at the photosphere (Fig. \ref{sketch} panel b). During the bald patch reconnection cool material is ejected with more   or less  fast  flows as we have shown in the explanation of the Mg II profiles.
The wavelengths of these Fe II and Ni II lines are located  at around  0.5  to 1 \AA\ far from the Si IV line center. Therefore  only when the Si IV line profiles are broad enough with extended wings,  
these chromospheric lines  are shown
themselves in absorption. It has been already observed in the  IBs \citep{Peter2014} and IRIS UV bursts \citep{Yan2015}.

 Moreover at the time of the mini flare (02:04:28 UT) a narrow dip in the profile of Si IV at its
rest wavelength is  observed, 
 blended by Fe II line (Figure \ref{MgII_SiIV}).  It could be due  multi-component flows like in the IBs \citep{Peter2014}. If that would be the
case, then one would expect the line profile to
be composed of two or more Gaussians at different
Dopplershifts representing the different
flow components. However  it looks not to be  the case and  the dip is deeper for  the strongest Si IV line 1394 \AA, and the dip is always at the rest wavelength.  Therefore this dip in Si IV could  be the signature of 
 opacity effects as in the UV burst presented in \citet{Yan2015}.
 The former authors show similar profiles of Si IV  with self absorption and with chromospheric lines  visible as  absorption lines.  They explain  these profiles by  the  superposition of different structures, in the deep atmosphere the reconnection leads to a significantly enhanced brightness and width of Si IV,  the  light passes through the overlaying cool structures where Ni II leads to absorption. Higher up there is emission of Si IV in overlying cool loops which lead to narrow self-absorption of Si IV. 
 This scenario is possible in our observations.  

In the UV burst bilateral outflows 
and expelled clouds with super Alfv\'enic flows of the order 200 km s$^{-1}$ have been observed  in Mg II and Si IV (Sect. \ref{UV_burst}). These profiles are similar to those of IBs found by \citet{Peter2014,Grubecka2016}. 
However,  we detected also O IV lines in the vicinity of  Si IV 1393.76 \AA\ (Table \ref{table2} and Fig. \ref{SiIV_profiles_slit4}). O IV are forbidden lines formed just below 0.2 MK.  Our UV burst is definitively not exactly an IB  where O IV lines were not detected   \citep{Peter2014} and  did not support the long  debate about the  temperature of the formation on Si IV line.
 Si IV could be  out of ionization equilibrium  in high velocity flow plasma and the nominal formation temperature of Si IV could  be in fact lower  than 80,000 K \citep{Dudik2014,Nobrega2018}. However when we observe simultaneously   O IV line emission  as well as Si IV  which is also  formed at transition region temperatures it confirms that the plasma  is heated and Si IV is not at  chromospheric temperature. In fact AIA observations showed the  mini flare  in its hotter filters until 10$^7$ K with 211 \AA\, filter.
 

\subsection{Optical thickness and the electron density} 	

{ Using  IRIS transition region lines (O IV, S IV and Si IV) electron density may  be computed \citep{Dudik2017,Polito2016,Young2018b}. We note that in the spectra corresponding to the reconnection site at the reconnection time, O IV lines are 
detected, even the emission is relatively weak (Figs. \ref{SiIV_evolution}, \ref{gradient_spectra_Mg} top panels and  \ref{SiIV_profiles_slit4}). Si IV line profiles vary drastically according to  time or location as we have already mentioned in Sect. \ref{UV_burst}.  
The great variety of shapes of profiles of the  Si IV lines 
rises a question about the variations of the optical thickness throughout the observed mini-flare area. For this investigation, we employ the  method involving 
the intensity ratio of the \ion{Si}{IV} 1393.75\,\AA~and 1402.77\,\AA~resonance lines \citep{Delzanna2002,Kerr2019}.
Since a long time this 
technique  exists  for stars  to determine the amount of opacity in the Si IV lines and is  very powerful 
 for providing the physical dimensions of the scattering  layer 
\citep{Mathioudakis1999}.

 For computing the intensity Si IV ratio   we select two observing times
: one time  during reconnection (02:04:28 UT) with slit 1 at the reconnection location (Fig. \ref{SiIV_evolution}), and the other  time at one minute later (02:05:36 UT) in slit 4 at the jet base (6\arcsec away  the reconnection point).}


\subsubsection{
Si IV integrated intensity}	
During the reconnection phase in slit position  1  at 02:04:28 UT, the extended profiles of the \ion{Si}{IV} resonance lines were obtained by summing the intensities observed at the different wavelengths throughout the profiles (Fig.\ref{SiIV_evolution}). 
We note that the absorption features were excluded from the analysis.
However the Si IV  1402 \AA\  are so wide (4 \AA),  overlying other transition region lines (O IV lines). We are not able to remove the contribution of these lines from  the integrated Si IV line intensity values. Based on the relatively-low intensities of these blending lines, we do not expect the uncertainty of this method to exceed 10\%.

Away from the reconnection region, in slit position 4 at 02:05:36 UT, we  compute the line integrated intensities  by fitting the observed profiles with Gaussian functions 
using the \texttt{xcfit.pro} fitting routine.
Typically, two Gaussian functions were needed to fit each line profile  of Si IV, in the mini-flare area 
between y = 76 and 88.  The two   Gaussian functions   
are separated by up to $\approx$0.4\,\AA, one function has  a narrow Full Width half maximum (FWHM)  and the other one  an extended  FWHM but with a lower intensity \citep{Dudik2017}. The  two Si IV line profiles showed
similar asymmetries. 
Particularly exceptional profiles of both lines were observed at y $\approx$80, where the `bumps' in line red wings were considerably weaker. Based on this, we suggest that the bumps present in the profiles do not originate in blends, but in the motion of the emitting plasma. To check  this assumption, we calculated synthetic spectra using CHIANTI v7.1 \citep{Dere1997, Landi2013} for log($N_{\text{e}}$ [cm$^{-3}$]) = 11 (see below), flare DEM. Even though we found several lines blending both lines of {Si} {IV}, their contributions were found to be negligible.

\subsubsection{Si IV ratio}

\begin{figure}[h]
  \centering    
    \includegraphics[width=0.48\textwidth]{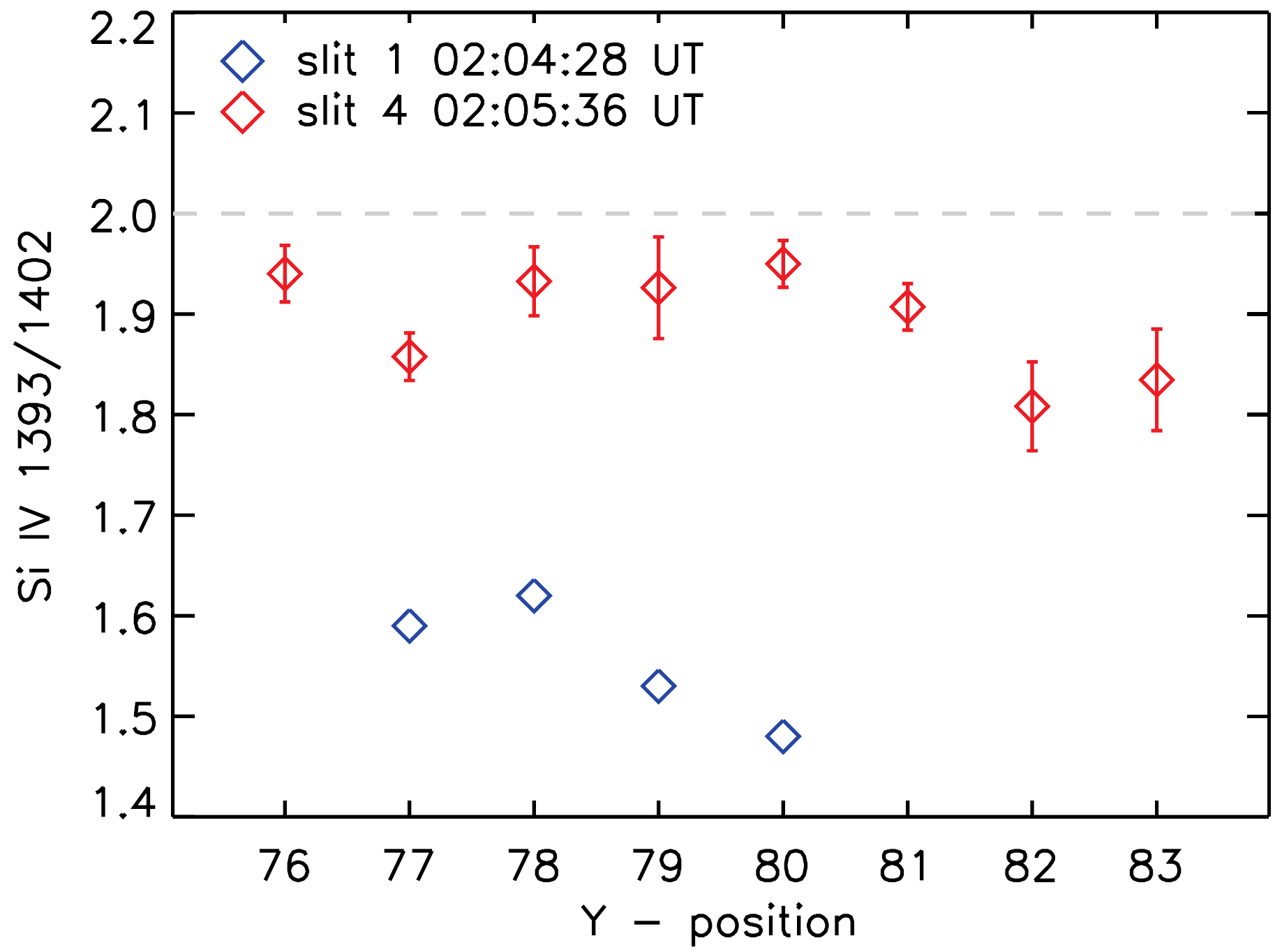}
  \caption{Variations in the Si IV 1393.75\,\AA~and 1402.77\,\AA\ line ratio used as a probe for measuring the optical thickness. Blue diamonds indicate the ratios measured in the reconnection region, while red ones away from it and later on. \label{fig_si_ratios}}
\end{figure}

The ratios of the \ion{Si}{IV} 1393.75\,\AA~and 1402.77\,\AA~resonance lines are shown in Fig. \ref{fig_si_ratios}. There, the diamonds indicate the measured ratios at different positions along the slit. They were color-coded in order to distinguish between the ratios measured in (blue) and away (red) from the reconnection region.

For optically thin plasma, this ratio should be equal  to 2 \citep{Delzanna2002, Kerr2019}, which we indicated using grey dashed line in Fig. \ref{fig_si_ratios}. Even though all of the observed ratios are below this value, the ratios measured farther from the reconnection region (red diamonds) are consistently closer to 2 than those measured in the reconnection region (blue diamonds).

The computed  ratio range for the red points,
  away of the reconnection site  is  between $\approx$1.72 and 1.95. The   maximum  values  are  around 1.9  for points 79 and 80 which infers to us  the  ability to derive the electron density.
  
For the blue points  in the mini-flare area at the reconnection time,   the low ratio  value of Si IV lines (1.48 to 1.62) with a high uncertainty  does not allow us to compute  the electron density 
  \citep{Kerr2019}. 
 More-less the Si IV line profile shapes are similar to the Mg II line shapes (Fig. \ref{MgII_SiIV})  with   similar extended blue wings which means that such large profiles has several components like the Mg II lines containing  certainly the emission of the flare plus the emission of a cloud with high velocities as we have concluded by analysing the Mg II profiles with the cloud model. It is nearly impossible to distinguish the two components in each Si IV profile and therefore to compute  the optical thickness of the cloud and the flare region.  
  The behaviour of Si IV line in flares  is exactly in a similar way as predicted in the theoretical models \citep{Kerr2019}, that there is a stratification in the heights at which the various lines  (Si IV, C II, Mg II) form, which varies with time in the flare. Initially the core of the Si IV 1393.75 line forms highest in altitude. Toward the end of the heating phase the compression of the chromosphere results as the lines formation in a very narrow region, which persists into the cooling phase. We believe to this stratification of heights of formation of Si IV, and C II lines during the flare. As it is
  suggested 
  by \citet{Kerr2019} the Si IV intensity ratio represents the ratio of the source function of the lines, the magnitude of the lines depending on the temperature of the layer where they are formed.  The thermalization of   both Si IV  lines  would occur higher in the atmosphere where the temperature is larger.  Tests using the RADYN  code should be used to understand such low ratio values in term of opacity effect. 
  One minute later after the flare this effect is negligible.


\subsubsection{Diagnostics of the electron density}
\begin{figure}
\centering
\includegraphics[width=0.48\textwidth]{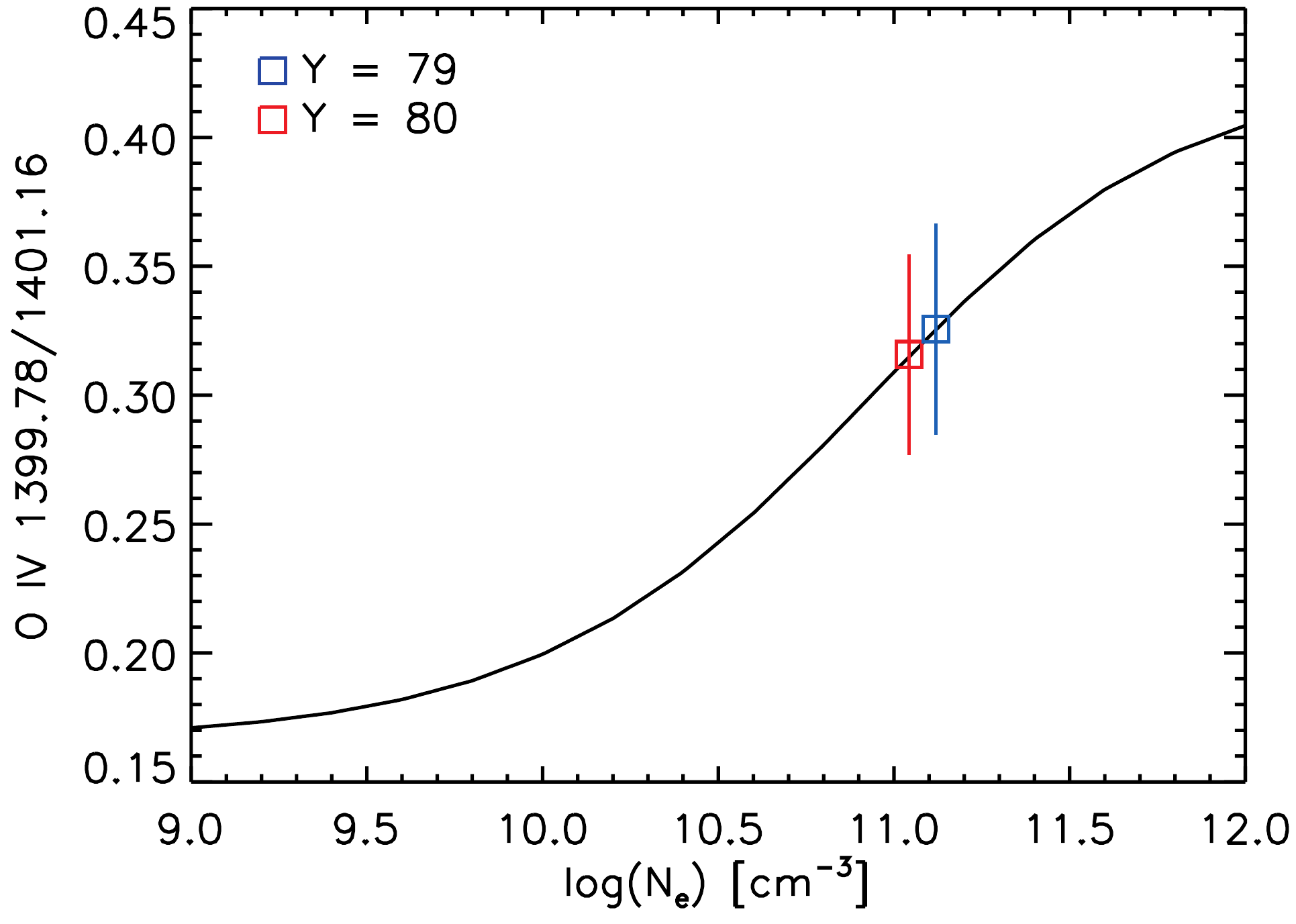}
  \caption{Diagnostics of the electron density using the intensities of the intercombination lines of O IV measured at y = 79 (blue), 80 (red). \label{fig_density}}
\end{figure}

The electron density could be computed in the mini flare area, away from the reconnection region (in slit 4) and observed a minute after the reconnection itself.
In order to obtain the electron densities 
we used density-sensitive ratios of line intensities. IRIS routinely observes multiple inter-combination lines which provide ratios useful for this purpose as they are not affected by opacity issues \citep{Polito2016,Young2018b}. Here, we are able to obtain reliable fits of lines composing only  one ratio, being the \ion{O}{IV} 1399.78\,\AA~and 1401.16\,\AA~lines. The former line was however still weak, with reliable fits only at slit 4 observed at 02:05:36 UT and  for y = 79 and 80.

The measured ratios are presented in Fig. \ref{fig_density}. The emissivities are calculated using the \texttt{dens\_plotter.pro} routine contained within the CHIANTI package for the peak temperature of formation of the ion (log($T$ [K]) = 5.15). At both locations, the resulting densities are roughly log($N_{\text{e}}$ [cm$^{-3}$]) = 11 $\pm$ 0.3. Note that these values correspond to those obtained by \citet{Polito2016} in a plage and a bright point. The mixed ratio proposed by \citet{Young2018b} with Si IV and O IV lines could not be computed due to the low counts.

\subsubsection{Path length}

Measured electron densities can then be used to calculate $\tau_{\text{0}}$ at the center of the line using e.g. Equation (21) of \citet{Dudik2017}:
\begin{equation}
	\tau_{\text{0}} \approx 0.26f \frac{\left\langle N_{\text{e}} \right\rangle}{10^{10} \text{ cm}^{-3}}, 
\end{equation}
where $f$ is the path length ($\triangle s$) filling factor in an $\textit{IRIS}$ pixel defined as $f = \triangle s/0.33 \arcsec$. The numerical factor of $0.26$  was by \citet{Dudik2017} calculated using thermal line widths of the resonance lines of \ion{Si}{IV} for the Maxwellian distribution. However, the observed profiles of the \ion{Si}{IV} lines are much broader. The widths resulting from fits are typically $>$0.5\,\AA, which reduces the numerical factor down to 0.012. Still, $\tau_{\text{0}}$ cannot be calculated unless the value of $f$ is known. If we for simplicity assume $f=1$ and utilize the measured density of log($N_{\text{e}}$ [cm$^{-3}$]) = 11, the modified formula leads to $\tau_{\text{0}} \approx 10^{-1}$. Note that at the same time and positions, ratios of the resonance lines were consistently closer to 2 compared to the values measured in the reconnection region, indicating relatively optically-thinner plasma. 

The formula for $\tau_{\text{0}}$ can easily be rewritten in terms of the path length $\triangle s$. Using the relation for $f$ and measurements of the line widths and electron densities, we obtain (in kilometres):
\begin{equation}
	\triangle s \approx 2000 \tau_{\text{0}}.
\end{equation}
Since we relaxed the assumption of $f=1$, the estimate for $\tau_{\text{0}}$ does not hold any further. However, as indicated by the resonance line ratios, outside of the reconnection region it is most likely $<1$ (Section 4.2.2). The possible path lengths are thus determined using the derived linear relation, while their upper boundary is $\approx$2000 km.

As it has been demonstrated for flares in stars \citep{Mathioudakis1999}
if the electron density in the atmosphere is
known, opacity can provide important information on the linear
dimensions of the scattering layer.

\begin{figure}
\centering
\includegraphics[width=0.46 \textwidth]{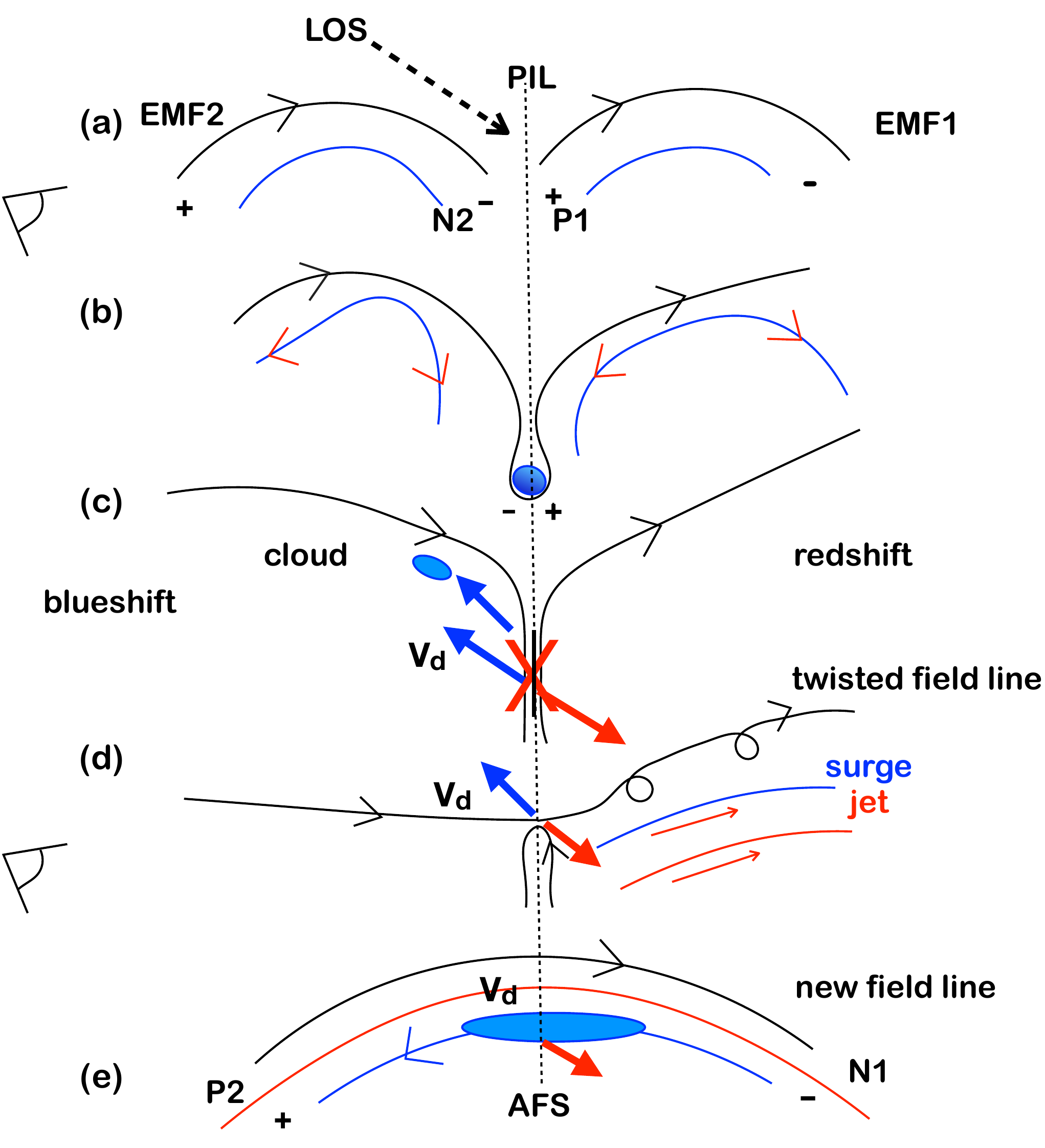}
\caption{Sketch of the dynamics of the plasma  in 2D (x,z) plane at the reconnection site before (panels a and b), during (panels c, d) and after  the magnetic reconnection (panel e). The sketch is based on the characteristics of the spectra detailed in table \ref{table4}. Panel (a) corresponds to the time 01:51 UT, panel (b) to 01:59 UT, panel (c) between 02:02:21 and 02:04:28 UT, panel (d) to 02:05:39 UT, and panel (e) to 02:10 UT. The eye on the left side indicates that the observations are  done on the side because the AR is located at W 60. The projected line of sight in the plane (x,z) is also shown by dashed arrow at the top. The magnetic field lines are shown with black solid lines  with arrows. The vertical central dotted line is the polarity inversion line (PIL) between magnetic polarities N2 and P1. The red solid lines are for hot jet (in panel d) or loops (in panel e) and the velocity directions are indicated by thin blue/red arrows for the transverse flow (panels b and e) and thick arrows for blue/redshifts  (V$_d$) estimated at the PIL zone (panels c, d, e) . In panel (a) EMF1 and EMF2 are two emerging magnetic flux (EMF), and 
N2-P1 is the bipole where the reconnection occurs in a bald patch with cool material trapped inside (blue circle in panel b). The blue lines and ovals are for the cool plasma visible with IRIS. The reconnection occurs at the `X' point with bidirectional outflows (panel c). The cloud is  a  blue oval on the left.
Nearly at  the same time  the jet (red line, hot plasma  visible in AIA filters) and the surge (blue line) are expelled  with some twist to the right (panel d). Panel (e) shows long loops (red line) and an arch filament system (AFS) (blue oval) between P2 and N1  after the reconnection.}
\label{sketch}
\end{figure}
\section{Discussion and Conclusion} \label{dis}
  A question arises concerning the height  of the magnetic reconnection of the mini flare at the jet base. When jets are observed over the limb like in the paper of \citet{Joshi2020},  the reconnection point  is clearly visible in the corona (e.g 10 Mm in the case of \citet{Joshi2020}). For events occurring on the disk it is difficult to derive the altitude of reconnection.  \citet{Grubecka2016,Reid2017} and  \citet{Vissers2019} used a NLTE radiative transfer code in a 1D atmosphere model to derive the altitude of formation of the  reconnection in UV burst visible in Mg II lines. The altitude range spans the high photosphere and chromosphere (50 to 900 km).  With IRIS it was also found  in UV bursts or IRIS bombs (IBs) that cool plasma  emitting in  chromospheric lines could overlay  hotter plasma at  Si IV  line temperature \citep{Peter2014}. After  a long debate about the temperature of the Si IV lines in possibly non equilibrium state,  3D simulation based on Bifrost code \citep{Gudiksen2011} coupled with the MULTI3D code \citep{Leenaarts2009} succeeded to mimic the large observed Si IV profiles (very similar to our UV  Si IV burst) and the  extended wing of the Ca K line with synthetic profiles both formed at different altitudes simultaneously due to an extended vertical current sheet  in  a strong magnetized atmosphere \citep{Hansteen2019}.  They proposed that  ``the current sheet is located in a large bubble of emerging
magnetic field, carrying with it cool gas from the photosphere". 
This scenario is certainly valid for our observations. 
We attempted  to compute the optical thickness and the electron density in the mini flare area during the reconnection process  \citep{Kerr2019,Judge2015}. The transition lines  (Si IV and O IV) usually used for such estimations had so perturbed profiles that no reliable numbers could be put forward at the time of the reconnection. One minute later the Si IV lines were estimated  to  be  nearly  optical thin  and the computed electron density arises to log($N_{\text{e}}$ [cm$^{-3}$]) = 11 $\pm$ 0.3,
 value which corresponds to a plage or bright point, similarly in the paper of \citet{Polito2016}.
 
\begin{figure}
\centering
\includegraphics[width=0.48 \textwidth]{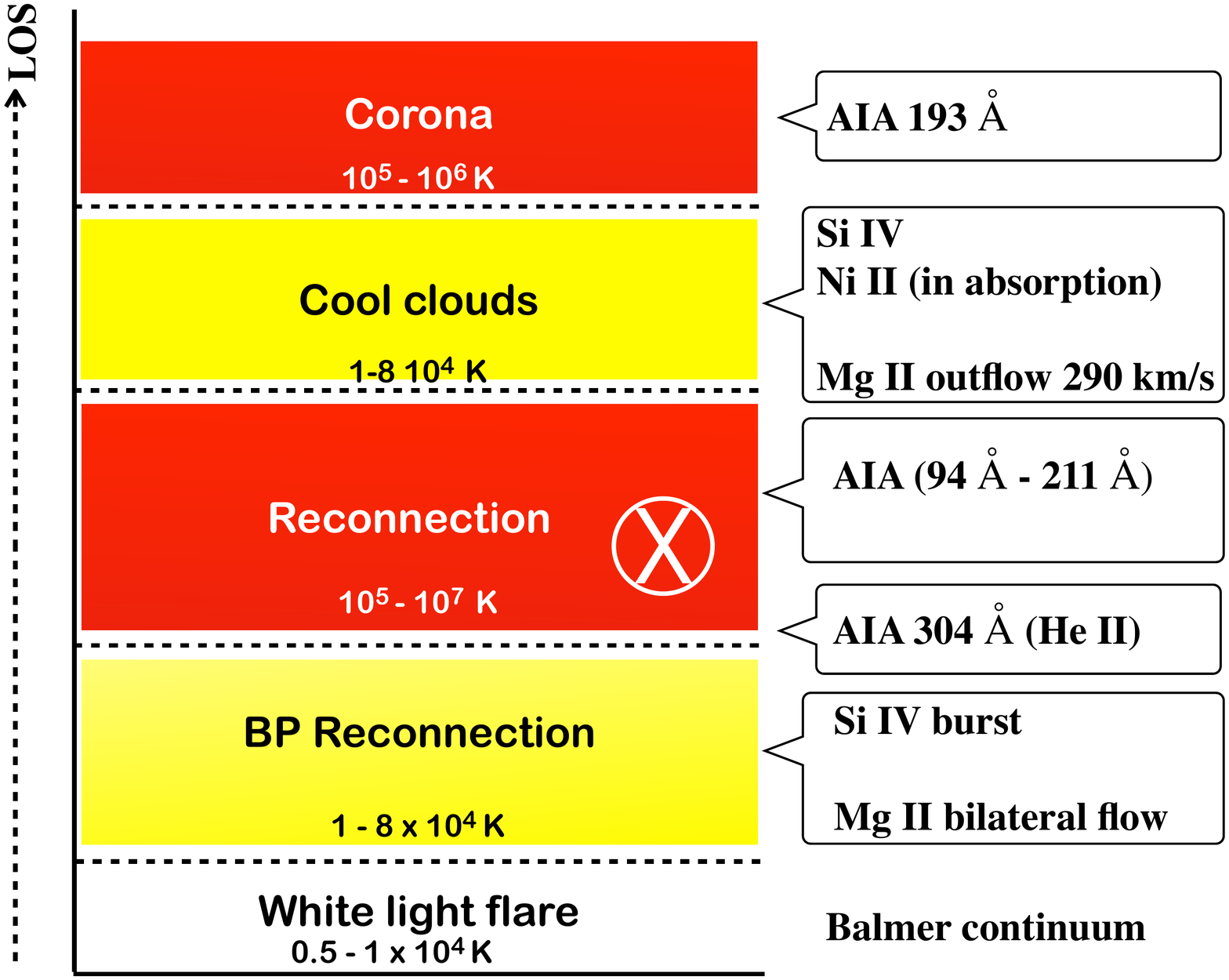}
\caption{Model of multi-layers of the mini flare atmosphere during the jet  reconnection in a bald patch region. The model is  based on the observations of emission or absorption of the IRIS lines and continua, and the images of AIA in the multi temperature filters in the mini flare (UV burst) around 02:04:28 UT $\pm$ 60 sec (right column).  The y axis is along the LOS.
The LOS crosses successively cool and hot layers  (white for minimum of temperature, yellow for chromosphere until transition region temperatures, red for coronal temperatures).
}
\label{sandwich}
\end{figure}

However, we have to consider that we have not only the signatures of IBs with IRIS but also
the
enhancement of  Balmer continuum detected in   IRIS spectra like in a white light mini flare.
The mini flare is  also visible as    brightenings  
in all  AIA filters with multi temperatures from 10$^5$ K to 10$^7$ K.
This mini flare, called mini flare due to its small area and GOES strength flux (B6.7)   is in fact a  very energetic  flare belonging to the category of white light flares. The chance is that  the slit of IRIS with its high spatial and temporal resolution was exactly at the site of the reconnection. Therefore it is the first time that we have such important information on a white light   mini flare.

We have all along Sect. \ref{Dopplershift} discussed on the signatures of the different elements and  proposed a dynamical model for the magnetic reconnection in Sect. \ref{multi-temps}. The analysis of HMI magnetogram of the AR shows that the region consists of several EMFs. The jet occurs between two of them when the negative polarity of the following EMF was collapsing with the  positive polarity of the leading EMF. Such magnetic topology  leads to  
a bald patch magnetic configuration 
where the magnetic field lines are tangent to the photosphere  (Paper I).
This topology was confirmed by the first reconnection  signature in  Mg II lines with symmetric extended wings at 02:03:46 UT (Fig. \ref{threeProfiles} panel (b)) similar to  those in IBs 
commonly occurring in bald patches \citep{Georgoulis2002,Zhao2017}.  At the reconnection site  these bilateral outflows ($\pm$ 200 km s$^{-1}$) observed in a few pixels were  interpreted as reconnection jet (Sect. \ref{UV_burst}). Less than one minute later (02:04:28 UT) extended Mg II line blue wing suggests super Alfv\'enic flows  (Fig. \ref{threeProfiles} panel (c).  With the cloud model technique, which represents a formal solution of the transfer equation under some assumptions, two plasma clouds, one with high speeds  (blueshifts of 290 km s$^{-1}$) and one with medium speed (blueshifts of 36 km s$^{-1}$) were detected (Sect. \ref{sec:cloud}). The identification of "explosive" i.e. 290 km s$^{-1}$  flow  is unique in such circumstance of reconnection.  We conjecture that cool plasma was trapped between the two EMFs  which could correspond to arch filament plasma and was expelled during the reconnection. The second cloud with lower velocity is certainly due to the surge plasma accompanying the jet. In the AIA 304 \AA\ movie, dark area was  observed at this time looking stationary. The surge plasma could extend along the LOS in a first phase before being elongated along the jet towards the West direction  at 02:05:39 UT (Fig. \ref{sketch} panel (d)).

\begin{table*}\caption{
Characteristics of the  evolution of the pattern and  Mg II spectra of the mini flare at the    jet  base (reconnection site in X) between the two arch filament systems observed in AIA 304 \AA\, and  in the  IRIS slit positions (1-4) (Figs. \ref{spectra} and  \ref{AIA_IRISspectra}). 
The  pixel numbers are the  coordinates  along the slit. This table is discussed in Sect. \ref{sec:sketch} and used to provide a  dynamical model of the reconnection  in Fig. \ref{sketch}.}
\bigskip
\centering
\resizebox{\textwidth}{!}{
\begin{tabular} {ccccc}
\hline

Time (UT) & AIA 304 \AA\ & \multicolumn{3}{c}{IRIS Mg II spectra} \\

& & (slit 1)& (slit 2-3)& (slit 4)\\

\hline
01 :51 :15 & Arch Filament Systems (AFS) & redshift at pixel 80 & high peaks & blue shift at pixel 80\\
       & over EMF1 and EMF2 & strong  central absorption & central absorption & central absorption\\
& & & &  \\
01 :54 :33& bright threads in X & blue/red shift at pixel 80 & broad lines& \\
& between the 2 AFS & central absorption & &\\
& & & &  \\
01 :56 :26& AFS with bright ends &long wings red and blue& large central absorption& thin blue\\
& & & &  \\
01 :59 :34& bright  threads in X  &ten y pixels with broad& \\
 & mixed with dark kernels &central absorption & & \\
 & & & &  \\
02 :01:10 & preflare in X & bright blue along 20 pixels & broad central absorption& thin \\
& & & &  \\
02 :02 :21 & bright NS arch &extended blue  wing  profile &blue  and broad & thin\\
& & & &  \\
02: 02: 59 & onset of surge&strong  central absorption& shift towards blue& \\
& & & &  \\
02 :03 :32 & kernels & very bright peaks symetrical & blue& blue \\
& & & &  \\
02 :04 :28& bright triangle jet base & bright  tilt blue to red  & large bright blueshift  & \\
& & & &  \\
02 :05 :39 & mini flare  & very bright blue tilt  & bright tilt &bright tilt\\
&double jet and surge&extended blueshift & & \\
& & & &  \\
02 :06 :22& expansion &zigzag central absorption & zigzag & elongated thin wing \\
& & & &  \\
02 :07 :04&  long surge over the jet&weak blue peak,  redshift& weak broad profiles & \\
& & & &  \\
02 :10 :21 & long loops& thin blue pixel& broad weak& continuum\\
 & bright and dark& weak extended redshift& & \\
 & & & &  \\
02 :14 :24 & long loops & one pixel extended  peaks & weak broad profiles& extended  thin blue-red  \\
   & long AFS & long red wing extension in y & red &\\
  & & & &  \\
02 :18 :24 & end  &  symmetric profiles& weak broad profiles& continuum \\
\hline
\label{table4}

\end{tabular} 
}
\end{table*}
 During the reconnection phase at 02:04:28 UT (Fig. \ref{threeProfiles} panels (b, h, m)) and Fig. \ref{sketch} (panel c)), we have many signatures of the different elements that allow us to build  a multi layer model  of the flare atmosphere along the current sheet.  We propose  for this flare  atmosphere the sandwich model with successive cool and hot plasma along the LOS (Fig. \ref{sandwich}).
Emission at the minimum temperature is  also detected with AIA 1600 \AA\ and 1700 \AA\ filters and confirmed this low level heating.
Mg II  and  C II  lines are good diagnostics for detecting plasma at chromospheric temperature (T $<$ 20,000K). We identified the bilateral outflows in the bald patches  still visible at this time of reconnection  in Si IV for example in Fig. \ref{SiIV_evolution}. The bald patch current sheet is  transformed to an `X'-point current sheet during the reconnection, 
which is responsible of the hot plasma detected in  the AIA filters (94 \AA\, to 211 \AA), as the MHD model suggests in Paper I.  However we have again the signature of cool plasma.  In fact 
Si IV  profiles   are  striped by  absorption lines formed  at  photospheric  temperatures and clearly show the presence of cool plasma 
over transition region temperature material in the  reconnection site (Sect. \ref{cool_over} and Fig. \ref{MgII_SiIV}). This cool plasma is certainly due to the cool clouds, either the ejected fast cloud  feed  by trapped material in bald patch or by surge plasma.
All this event is  finally embedded  in the corona.
This  demonstrated  the possibility of having successive layers in the atmosphere with different velocities and temperatures in the current sheet region.  
 
We have used the cloud model  technique as a simple diagnostics tool. Such models are used to derive  true velocities which usually differ from
those obtained from Doppler shifts (this is the basic idea behind the cloud
model). Then the question is what is the nature of  the obtained velocities,
 expelled  plasma blobs like in surges \citep{Moreno2008,Nobrega2018} as we suggested, upflows (evaporation), downflows (chromospheric condensations) as it is proposed for flares \citep{Berlicki2005,DelZanna2006}. However 
using the cloud-model technique cannot help us for the 
proper understanding
of the nature of detected flows, it is just a diagnostics method. Numerical simulations using RADYN and RH \citep{Kerr2019} or   Flarix RHD codes \citep{Kasparova2019}  would certainly give more insights  on the physical process involved. These observations could be the boundary conditions of future simulations. As we have discussed in section 4.2 we  found similarities between such simulation models and our  observations like  the high variability of the Si IV lines depending strongly on opacity effects.

\begin{figure}[ht!]
\centering
\includegraphics[width=0.48 \textwidth]{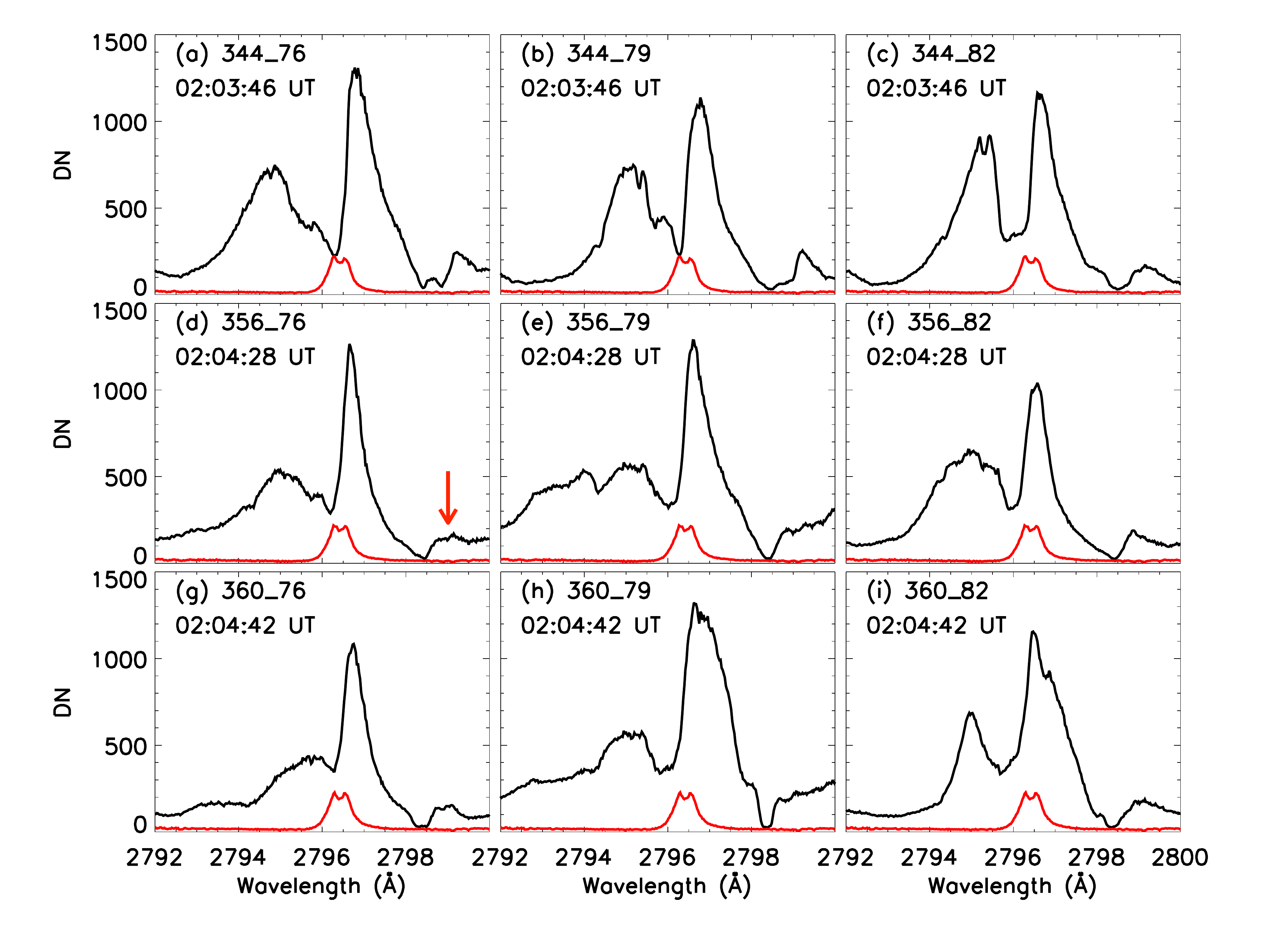}
\caption{Evolution of Mg II k line profiles  inside the UV burst for three times (each row distant of 42 s and 14 s respectively) and for three pixels  distant of one  arcsec  (between each column).The UV burst in located in slit 1.  The red profile is the reference profile in the quiet chromosphere. The red arrows  indicate the emission of the Mg II triplet lines. \label{burst_MgII}}
\end{figure}
\begin{figure}[ht!]
\centering
\includegraphics[width=0.48 \textwidth]{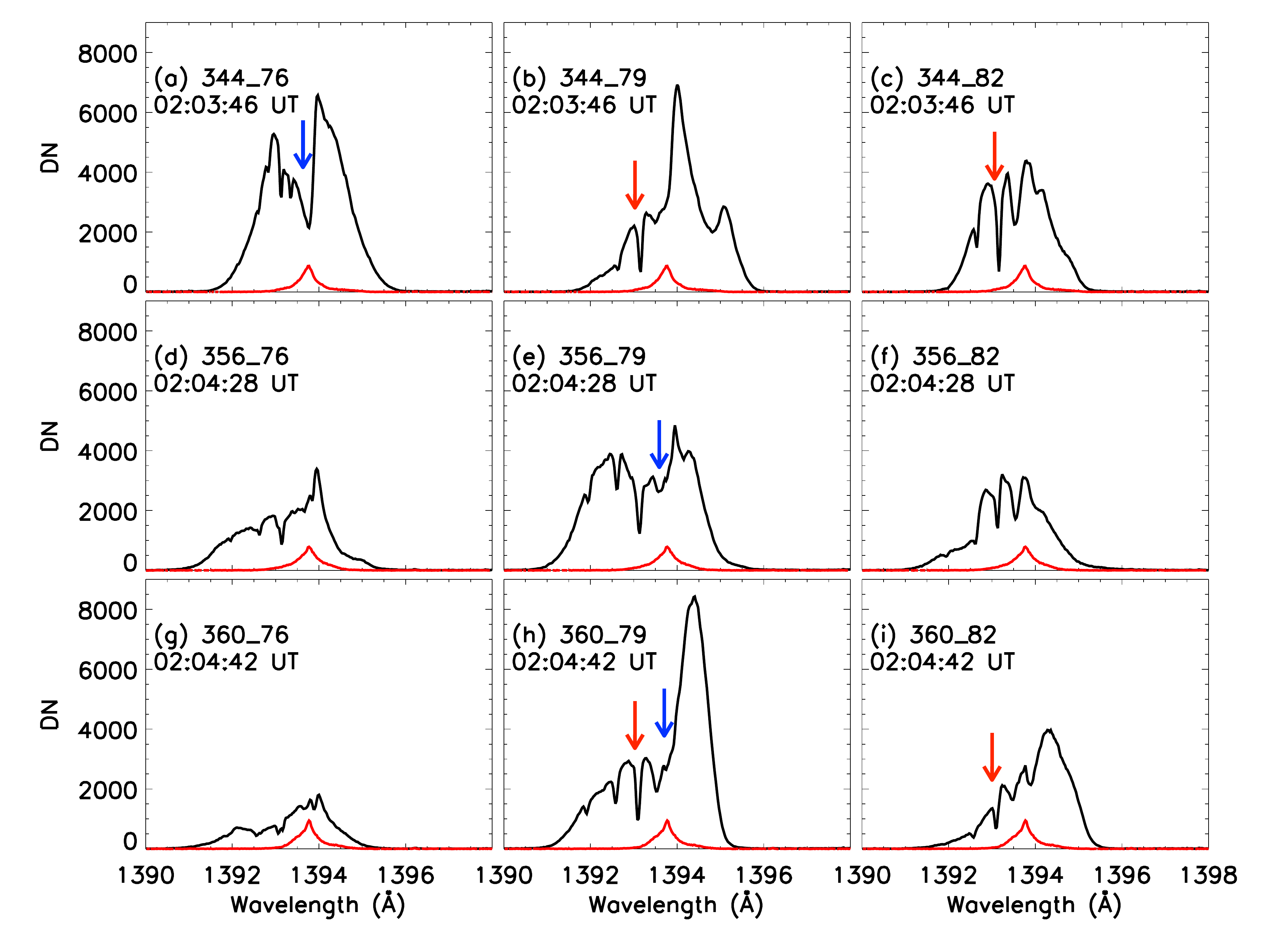}
\caption{Evolution of  Si IV 1393.76 \AA\ line profiles  inside the UV burst (same points as in Figure \ref{burst_MgII}) for three times (each row distant of 42 s and 
14 s respectively) and for three pixels  distant of one  arcsec  (between each column).The UV burst in located in slit 1.  The red profile is the reference profile in the quiet chromosphere.  The red arrows indicate the absorption by the  Ni II 1393.33 \AA\, and Fe II 1393.589 \AA.  The blue arrows indicate the dip in the central Si IV profile due to self absorption.\label{burst_SiIV}}
\end{figure}
\begin{center}
ANNEX
\end{center}
The spatio-temporal variations of the Mg II k and Si IV profiles are displayed
for three times and for three different pixel values 
in Figures \ref{burst_MgII} and \ref{burst_SiIV}. The profiles vary fast on these time and spatial scales (in 30 sec. and 1$\arcsec$ respectively).  We note the need of high spatial and temporal resolution spectra to detect with accuracy the time and the location of reconnection.


\begin{acknowledgements}
We thank to the  anonymous referee for their valuable suggestions and comments. We would like to thank Dr Jaroslav Dudik for his fruitful discussions on the opacity of the Si IV lines. We thank the SDO/AIA, SDO/HMI, and IRIS science teams for granting free access to the data.  RJ acknowledges to CEFIPRA for a Raman Charpak Fellowship under which this work is done at Observatoire de Paris, Meudon. RJ thanks to Department of Science and Technology (DST), New Delhi, India for the INSPIRE fellowship. BS and GA acknowledge financial support from the Programme National Soleil Terre (PNST) of the CNRS/INSU. 
The work of JL is supported by the Charles University, project GA UK 1130218.
The work of RC is supported by the
Bulgarian Science Fund under Indo-Bulgarian bilateral project.PH and JL acknowledge support by Czech Funding Agency grant No. 19-09489S. Authors thank providers of open-source software for online
calls and meetings, which were essential for the completion
of this work during the outbreak of the COVID-19
pandemic.
\end{acknowledgements}

\bibliography{references-4}

\bibliographystyle{aa}
\end{document}